\documentclass[letterpaper]{article} 
\usepackage{aaai2026}  
\usepackage{times}  
\usepackage{helvet}  
\usepackage{courier}  
\usepackage[hyphens]{url}  
\usepackage{graphicx} 
\usepackage{natbib}  
\usepackage{caption} 
\usepackage{algorithm}
\usepackage{algorithmic}
\usepackage{enumitem}
\usepackage{amssymb}
\usepackage{mathtools}
\usepackage{amsthm}

\usepackage{amsmath}
\usepackage{transparent}
\usepackage{color}
\usepackage{xcolor}
\usepackage{colortbl}
\usepackage[makeroom]{cancel}
\usepackage{soul}  
\usepackage{pifont}
\usepackage{rotating} 
\usepackage{float}
\usepackage{booktabs}

\usepackage{amsmath,amsfonts,bm}

\def\eqref#1{equation~\ref{#1}}

\def\1{\bm{1}}

\def\vh{{\bm{h}}}

\def\mA{{\bm{A}}}
\def\mB{{\bm{B}}}
\def\mC{{\bm{C}}}
\def\mD{{\bm{D}}}

\DeclareMathAlphabet{\mathsfit}{\encodingdefault}{\sfdefault}{m}{sl}
\SetMathAlphabet{\mathsfit}{bold}{\encodingdefault}{\sfdefault}{bx}{n}

\usepackage{multirow} 

\definecolor{gray}{rgb}{0.46,0.46,0.46}
\definecolor{darkergreen}{RGB}{21, 152, 56}
\definecolor{darkerred}{RGB}{220, 35, 120}
\definecolor{darkerblue}{rgb}{0,0.08,0.45} 
\definecolor{royalblue}{RGB}{65,105,225}
\definecolor{lightblue}{RGB}{221,235,247}
\definecolor{gray94}{gray}{.94}
\definecolor{gray90}{gray}{.90}

\newcolumntype{g}{>{\columncolor{gray94}}c} 
\usepackage{newfloat}
\usepackage{listings}
\DeclareCaptionStyle{ruled}{labelfont=normalfont,labelsep=colon,strut=off} 
\floatstyle{ruled}
\newfloat{listing}{tb}{lst}{}
\floatname{listing}{Listing}
\title{TrinityDNA: A Bio-Inspired Foundational Model for Efficient \\ Long-Sequence DNA Modeling}

\author{
    Qirong Yang\textsuperscript{\rm 1}$^*$,
    Yucheng Guo\textsuperscript{\rm 1}$^*$,
    Zicheng Liu\textsuperscript{\rm 1,2}\thanks{First three authors contribute equally.},
    Yujie Yang\textsuperscript{\rm 1},
    Qijin Yin\textsuperscript{\rm 1},
    Siyuan Li\textsuperscript{\rm 1,2},\\
    Shaomin Ji\textsuperscript{\rm 1},
    Linlin Chao\textsuperscript{\rm 1},
    Xiaoming Zhang\textsuperscript{\rm 1}\thanks{Corrsponding author.}
}
\affiliations{

    \textsuperscript{\rm 1}BioMap Research, Beijing, China\\
    \textsuperscript{\rm 2}AI Lab, Research Center for Industries of the Future, Westlake University, China
}

\usepackage{bibentry}

\begin{document}

\maketitle

\begin{abstract}
The modeling of genomic sequences presents unique challenges due to their long length and structural complexity. Traditional sequence models struggle to capture long-range dependencies and biological features inherent in DNA. In this work, we propose TrinityDNA, a novel DNA foundational model designed to address these challenges. The model integrates biologically informed components, including Groove Fusion for capturing DNA's structural features and Gated Reverse Complement (GRC) to handle the inherent symmetry of DNA sequences. Additionally, we introduce a multi-scale attention mechanism that allows the model to attend to varying levels of sequence dependencies, and an evolutionary training strategy that progressively adapts the model to both prokaryotic and eukaryotic genomes. TrinityDNA provides a more accurate and efficient approach to genomic sequence modeling, offering significant improvements in gene function prediction, regulatory mechanism discovery, and other genomics applications. Our model bridges the gap between machine learning techniques and biological insights, paving the way for more effective analysis of genomic data. Additionally, we introduced a new DNA long-sequence CDS annotation benchmark to make evaluations more comprehensive and oriented toward practical applications.
\end{abstract}
\section{Introduction}
The rapid advancements in large-scale, long-sequence modeling, particularly in the realm of Natural Language Processing (NLP)~\citep{team2023gemini, achiam2023gpt}, have radically transformed the way we approach complex data. Deep learning models, such as Transformers~\citep{vaswani2017attention}, have achieved unprecedented success in tasks that span from language translation to text generation, revolutionizing not just NLP but a variety of other fields. These models have proven their capability to capture intricate dependencies in data, providing solutions to challenges that were once considered insurmountable. With these breakthroughs in NLP, there has emerged an exciting opportunity to extend the power of sequence modeling to a completely different domain—genomics—where data shares some key similarities, such as its sequential nature.

Genomic data, particularly DNA sequences, consists of extraordinarily long strings of information that encode the fundamental building blocks of life. Unlike the highly dense and structured data typically encountered in NLP, genomic sequences are sparse in nature, containing vast stretches of repetition and variability~\citep{liu2024genbench}. Despite this, they hold a rich repository of biological information that is crucial for understanding gene functions, regulatory mechanisms, and cellular processes. The ability to model DNA sequences deeply could lead to breakthrough applications in personalized medicine, genetic engineering, and the overall understanding of biological systems. However, effectively capturing the dependencies within such long, sparse sequences remains a significant challenge.

\begin{figure*}[t!]
    \centering
    \includegraphics[width=1.0\linewidth]{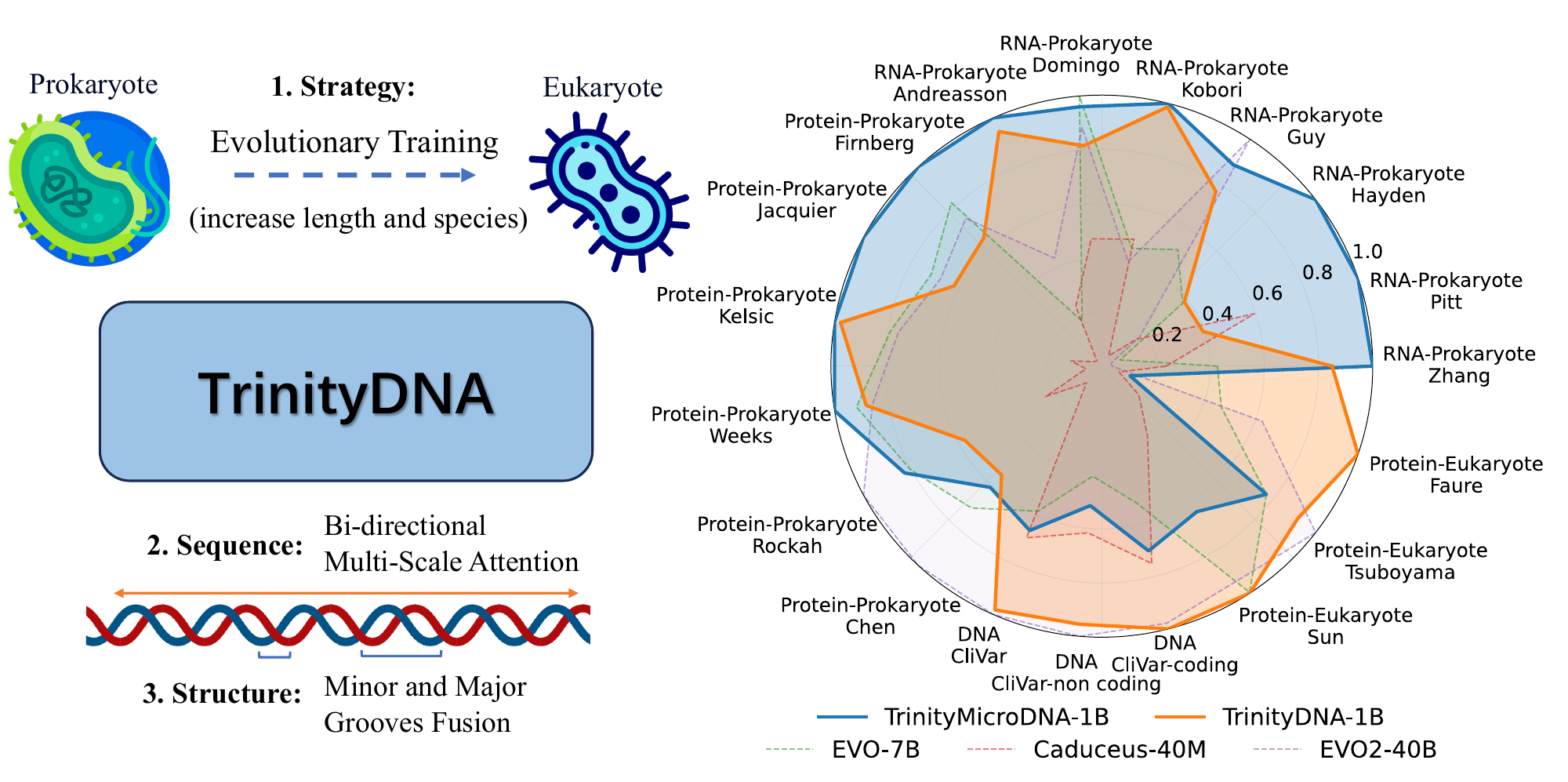}
    \caption{Overview of TrinityDNA Model: (Left) The evolutionary training strategy of TrinityDNA, progressing from prokaryotic DNA to multi-species eukaryotic DNA, and its DNA-targeted long-sequence modeling approach addressing structural features such as bidirectional complementarity and major/minor grooves. (Right) Radar chart illustrating the state-of-the-art performance on the zero-shot performance of our models versus popular models such as EVO and Caduceus.}
    \label{fig:intro}
\end{figure*}

While the parallels between NLP and genomics are evident, directly applying traditional NLP models to genomic sequences proves difficult. The sparse, low-density nature of DNA sequences means that existing models often struggle to identify long-range dependencies and interpret the underlying biological structures\citep{mallet2021reverse, zhou2021towards}. Moreover, the lack of biologically informed features in current models limits their effectiveness in genomic contexts. Many models trained on single-species data perform poorly when generalized to other species or broader biological contexts. As a result, the impact of these models on genomic research has been somewhat restricted, and their applicability to real-world challenges remains limited.

To address these issues, we introduce TrinityDNA, a novel DNA foundational model specifically designed to overcome the current limitations of genomic sequence modeling, as shown in Fig~\ref{fig:intro}. TrinityDNA leverages the latest advancements in deep learning to create a model that is optimized for the unique challenges posed by DNA sequences while also incorporating key biological insights. The contributions are listed as follows:
\begin{itemize}
    \item \textbf{Bio-inspired Design}: A multi-level architecture that is optimized for DNA sequences and leverages the Groove Fusion module and Reverse Complement (RC) fusion strategy. This design captures and exploits the unique structural properties of DNA, enabling long bi-directional genomic modeling.
    \item \textbf{Evolutionary Training Strategy}: A multi-species training regimen that spans a variety of organisms from prokaryotes to eukaryotes, enabling the model to generalize different genomic contexts and sequence lengths.
    \item \textbf{Comprehensive Large-Scale Data Integration}: Curated and integrated datasets from prominent genomic databases such as GTDB, IMG, and RefSeq, ensuring a diverse and high-quality foundation for model training.
    \item \textbf{New Benchmark for Long-Sequence Inference}: We introduce a novel \textit{CDS Annotation Benchmark} that focuses on gene-structure labeling in prokaryotic genomes, assessing both long-sequence modeling and practical annotation performance.
\end{itemize}


\section{Background}
\subsection{DNA Terminology for Structures.}
\paragraph{Basic Composition}
Deoxyribonucleic acid (DNA) is the hereditary material in most living organisms, consisting of long sequences of nucleotides. Each nucleotide comprises a phosphate group, a deoxyribose sugar, and one of four nitrogenous bases: adenine (A), thymine (T), cytosine (C), and guanine (G). A DNA sequence can be represented as a string \( S = (s_1, s_2, \ldots, s_N) \), where \( s_i \in \{A, T, C, G\} \) and \( N \) denotes the sequence length. 

\paragraph{Minor and Major Grooves.}
The double helix structure of DNA features two distinct grooves: the minor groove and the major groove. These grooves arise from the asymmetric positioning of the phosphate backbone relative to the base pairs. 
(1) \textit{Major Groove}: wider and deeper, the major groove provides greater accessibility for protein binding and molecular interactions. It generally covers five to seven nucleotides.
(2) \textit{Minor Groove}: narrower and shallower, the minor groove presents a different arrangement of hydrogen bond donors and acceptors. While less accessible than the major groove, its length is three to five nucleotides.

\paragraph{Reverse Complement Strands.}
DNA molecules consist of two complementary strands running in opposite directions, a feature known as antiparallel orientation. For a DNA sequence \( S = (s_1, s_2, \ldots, s_N) \), its reverse complement \( S^R \):
\[
S^R = (s_N^C, s_{N-1}^C, \ldots, s_1^C)
\]
where \( s_i^C \) denotes the complementary base of \( s_i \) following the base-pairing rules: \( A \leftrightarrow T \) and \( C \leftrightarrow G \).
This reverse complementarity is fundamental to DNA replication and transcription processes. Incorporating reverse complement information into computational models enhances their ability to capture symmetrical and complementary patterns, thereby improving predictions related to gene annotation and regulatory element identification.

\subsection{DNA Long-Sequence Modeling}
\label{sec:long_sequence_models}

\paragraph{Structured State Space Models (SSMs).}
\label{subsec:ssm}
A prominent class of models for handling long-range dependencies is based on \emph{Structured State Space Models (SSMs)}~\citep{gu2021efficiently, 
gu2021combining, gu2022parameterization, gupta2022diagonal, smith2022simplified,dao2022hungry}. These models emerge from discretizing a continuous-time 
linear system:
\[
\begin{aligned}
    {\vh}(t) &= \mA \vh(t) + \mB\,x(t), \quad &&y(t) = \mC\,\vh(t) + \mD\,x(t), \\
    \vh_{t+1} &= \overline{\mA}\,\vh_t + \overline{\mB}\,x_t, \quad 
    &&y_{t+1} = \mC\,\vh_t + \mD\,x_t,
\end{aligned}
\]
where \(\vh(t)\) (or \(\vh_t\)) is an internal state that can capture long-range dependencies in the input sequence \( x(t)\) (or \( x_t\)). Through efficient convolution-based implementations of these recurrences, SSMs have shown strong 
performance on very long sequences. However, conventional SSMs do not explicitly adapt their parameters to specific tokens or positions, which may limit expressivity for tasks such as DNA sequence modeling. Therefore, a line of SSMs are designed for long-sequence DNA.


\paragraph{HyenaDNA (EVO).}
\label{subsubsec:hyenadna}
\emph{HyenaDNA}~\citep{poli2023hyena} is an SSM variant, a decoder-style model capable 
of handling long genomic sequences (e.g., hundreds of thousands of tokens). 
It leverages the \emph{Hyena} operator, which replaces traditional attention with a 
fast convolution-based mechanism. Concretely, HyenaDNA blocks compute a Toeplitz 
convolution filter on projected input tokens, allowing processing of very 
large contexts (in \(\mathcal{O}(L \log L)\) time) without the quadratic cost typically 
associated with attention.

\paragraph{Caduceus (MambaDNA).}
\label{subsubsec:mambadna}
\emph{MambaDNA}~\citep{schiff2024caduceus}---also referred to as \emph{Caduceus}---builds on the selective SSM 
approach of Mamba and incorporates \emph{reverse-complement symmetry}, a core property 
of DNA sequences. In standard Mamba, the module processes sequences in a single 
direction. MambaDNA extends this design in two ways:
(1) \textbf{BiMamba}: Instead of only left-to-right processing, BiMamba applies 
    the Mamba block twice.
(2) \textbf{RC Equivariance}: MambaDNA explicitly enforces RC symmetry by taking 
    a sequence and its RC as inputs to the same SSM-based module.
See more details in Appendix~\ref{app:related}.


\section{TrinityDNA: Sequence, Structure, and Strategy for DNA Modeling}

\begin{figure}[b!]
  \begin{center}
    \includegraphics[width=0.95\linewidth]{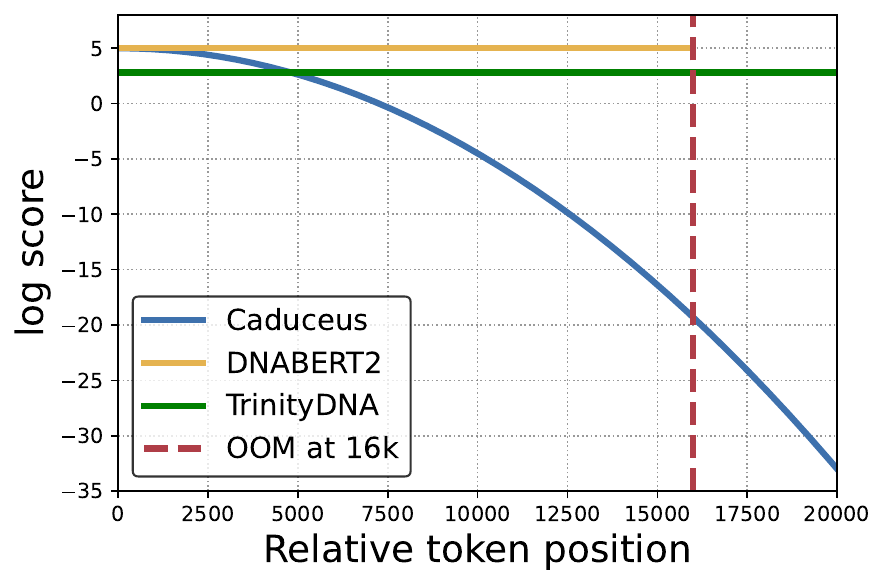}
  \end{center}
  \caption{Comparison of log influential scores log \(|\partial yt/\partial xs|\) versus distance \((t-s)\) on HG-38~\cite{nguyen2023hyenadna}.}
  \label{fig:pre}
\end{figure}

\subsection{Preliminaries}
\paragraph{Lost in the locality.}
While SSMs theoretically excel at handling long sequences, they inherently exhibit a \textit{locality bias}~\citep{wang2024understand}. This issue is exacerbated in DNA sequence modeling, where dependencies across vast genomic regions must be captured to enable accurate biological interpretation. Specifically, in genomic data, long-range dependencies often span tens of thousands or even hundreds of thousands of base pairs. Existing models, such as SSMs, typically focus on local dependencies due to computational challenges in processing long sequences. Consequently, our empirical results in Figure~\ref{fig:pre} demonstrate that the SSM-based model (Caduceus) lost their focus as the sequence length increased, while the full-attention-based model (DNABERT2) suffered from heavy computation. These limitations hinder modern DNA foundation models' ability to fully capture the inherent complexities of DNA sequences and their interconnections across genomic regions.

\begin{figure}
  \begin{center}
    \includegraphics[width=0.95\linewidth]{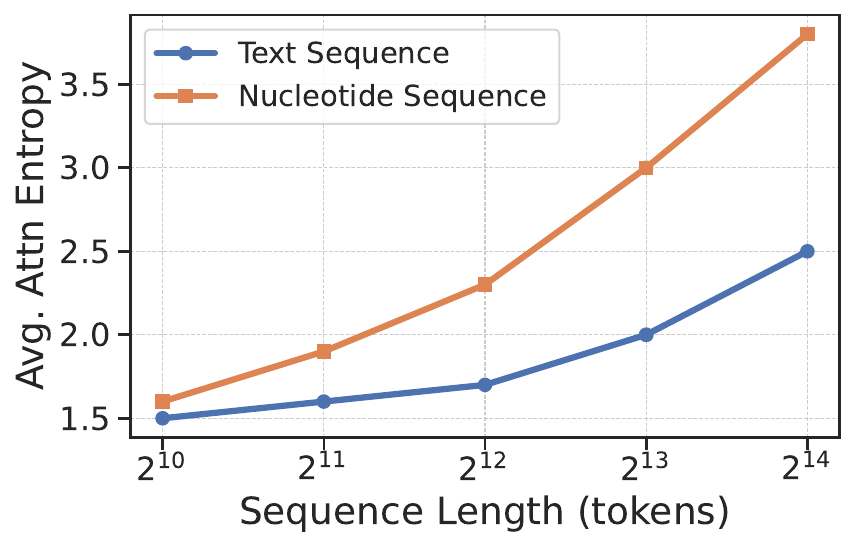}
  \end{center}
  \caption{Average attention entropy of full self-attention models as sequence length increases.}
  \label{fig:over}
\end{figure}
\paragraph{Oversmoothing in long attention.}
As sequences grow in training, full self-attention “flattens out”: attention scores converge toward a uniform, high-entropy distribution in Figure~\ref{fig:over}, so every token is weighted almost equally, and useful signals vanish. This oversmoothing affects low-density data—images, DNA, and the hardest— where informative tokens are sparse and far apart. In shorter windows, the same model still shows specialized heads, \textit{e.g.}, retrieval vs. induction~\citep{xiao2024duo}, but that diversity collapses at the kilobase scale. These findings motivate a multi-window, multi-head scheme that combines narrow local windows with broader global ones to reduce entropy and maintain head specialization. Detailed settings refers to the Appendix.

\begin{figure*}[t!]
    \centering
    \includegraphics[width=0.9\linewidth]{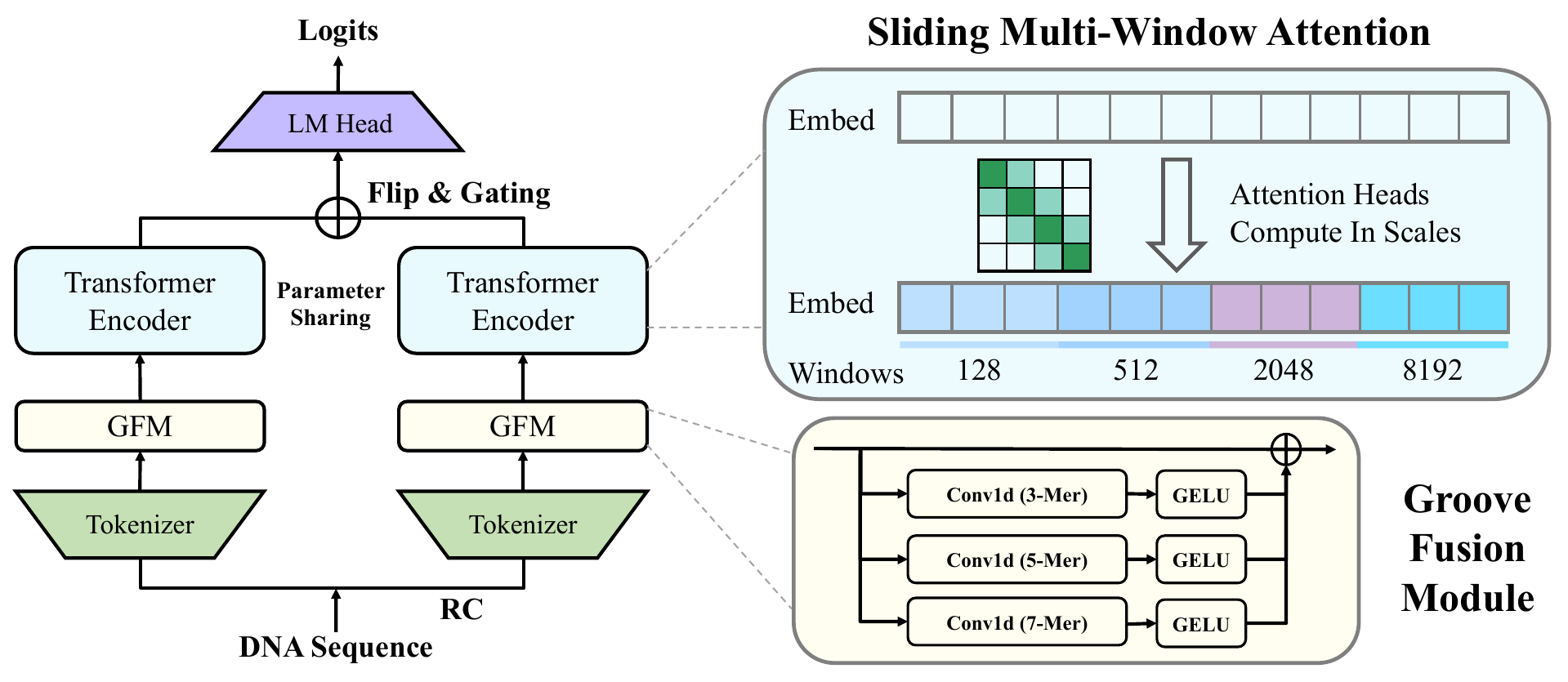}
    \caption{Model Architecture of TrinityDNA: The model integrates DNA sequences and structural features by considering its grooves and reverse complementary sequence with shared parameters.}
    \label{fig:model}
\end{figure*}

\subsection{Groove Fusion Module}
To account for the minor and major grooves in DNA sequences, we propose a Groove Fusion module that combines convolutional operations of varying kernel sizes. These grooves have distinct structural and functional roles in DNA, with the major groove offering greater accessibility for protein binding and the minor groove being involved in different molecular interactions. To model these differences, we perform tokenization on the DNA sequence using three convolutional kernels of sizes 3, 5, and 7. This multi-scale convolution approach enables the model to focus on different spatial features across the sequence, effectively capturing the structural nuances associated with the two grooves.

Formally, the Groove Fusion process can be defined as:
\[
\text{GrooveFusion}(S) = \sum_{k \in \{3, 5, 7\}} \text{GELU}(\text{Conv}_k(S))
\]
where \( \text{Conv}_k \) represents the convolution operation with kernel size \( k \), and \( S \) is the input DNA sequence. The output of each convolution operation is fused to capture the multi-scale contextual information necessary for interpreting the structural variations between the major and minor grooves.

\subsection{Sliding Multi-Window Attention}
To mitigate locality bias and attention oversmoothing, we revisit the design of Multi-Head Attention (MHA), focusing on varying attention window sizes across heads. In traditional MHA, each head typically attends to the entire sequence using a fixed attention window, which can limit the model's ability to capture dependencies at different scales. 
To address this, we introduce a multi-scale attention grouping strategy named SMWA. In this approach, we assign different attention heads to capture dependencies at different scales in the DNA sequence, enabling the model to specialize in local or global dependencies, depending on the feature scale.
Formally, we define the window sizes for each attention head \( h \) as \( L_h \in \mathbb{N} \), with \( L_h \) representing the length of the attention window for head \( h \). 
For head \( h \), the attention mechanism is computed using a sliding window of size \( L_h \) over the sequence, allowing the model to focus on scales:
\begin{align*}
    \text{Attn}_h(S_i) = & \text{Softmax}\left(\frac{Q_h(i) K_h(i + [-L_h, L_h])^T}{\sqrt{d_k}}\right) \\& V_h(i + [-L_h, L_h])
\end{align*}
where \( Q_h \in \mathbb{R}^{N \times d_k} \) is the query matrix for head \( h \), \( K_h \in \mathbb{R}^{N \times d_k} \) is the key matrix, \( V_h \in \mathbb{R}^{N \times d_v} \) is the value matrix, \( N \) is the sequence length, and \( d_k \) and \( d_v \) are the dimensionality of the key and value vectors, respectively. \( i \) is the index of the sequence, and \( [-L_h, L_h] \) represents the range of indices for the sliding window around \( i \). This enables each head to specialize in attending to either short-range or long-range dependencies by adjusting \( L_h \).
The final output of the multi-head attention layer is the concatenation of all heads' outputs, followed by a linear transformation:
\[
\text{SMWA}(S) = \text{Concat}(\text{Attn}_1, \text{Attn}_2, \dots, \text{Attn}_H) W_O
\]
where \( W_O \in \mathbb{R}^{d_{\text{out}} \times d_{\text{in}}} \) is a learned output weight matrix, \( H \) is the number of attention heads, and \( d_{\text{out}} \) is the output dimensionality.
This multi-scale attention mechanism allows the model to simultaneously capture both local and global dependencies by allocating different heads to focus on different sequence scales. Thus, shorter sequences can be captured by heads with smaller windows, while longer-range dependencies can be modeled by heads with larger windows, enabling the model to capture the hierarchical nature of DNA sequences better.

\subsection{Gated Reverse Complement}
\begin{figure*}
    \centering
    \includegraphics[width=0.85\linewidth]{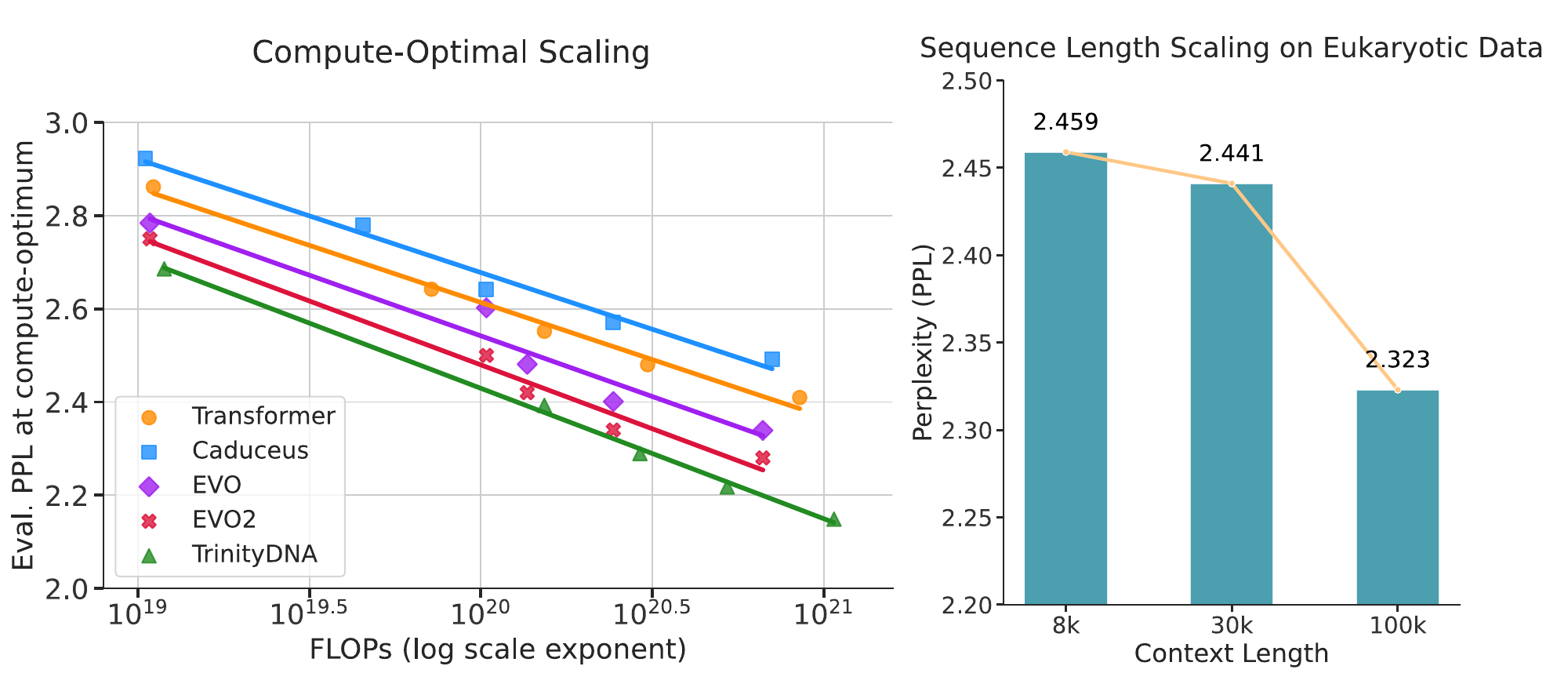}
    \caption{\textbf{Scaling Behaviors of Our Proposed Model.} 
    \emph{(Left)}~Evaluation perplexity (PPL) against total FLOPs across multiple architectures, showing consistent improvements to various baselines. 
    \emph{(Right)}~Impact of increasing context length (8k, 30k, 100k) on a eukaryotic dataset, where PPL steadily decreases with longer context windows.}
    \label{fig:scaling_law}
\end{figure*}
To leverage the reverse complementarity inherent in DNA sequences, we introduce a novel Gated Reverse Complement (GRC) mechanism. This mechanism employs a shared Transformer module to process both the forward and reverse complement sequences in parallel. The reverse complement of a DNA sequence \( S = (s_1, s_2, \dots, s_N) \) is defined as \( S^R = (s_N^C, s_{N-1}^C, \dots, s_1^C) \), where \( s_i^C \) denotes the complementary base of \( s_i \) (with base-pairing rules \( A \leftrightarrow T \), \( C \leftrightarrow G \)).
The GRC mechanism works by feeding both the forward and reverse complement sequences into a shared SMWA-equipped Transformer network \( f_\theta \), where the outputs are gated using a linear gating mechanism to combine the two representations effectively. The final layer is:
\begin{align*}
    \text{Output} = \text{GRC}(S, S^R) = & f_\theta(S) + \sigma(W_G \cdot f_\theta(\text{Flip}(S^{R})))
\end{align*}
where \( W_G \) are learned weights, \( \sigma \) represents the identity function, and the Flip operator means to reverse the sequence in the original order. This allows the model to learn the representations of both sides simultaneously.

\subsection{Evolutionary Training Strategy}

The Evolutionary Training Strategy (ETS) approach leverages a two-stage, evolution-inspired training strategy to progressively address the varying complexities of genomic data. In the first stage, the model is trained on prokaryotic genomes, which are relatively straightforward in terms of their regulatory architectures \citep{xu2006genome, nguyen2024sequence}. Through this foundational phase, the model captures essential DNA sequence motifs and organizational patterns. Subsequently, the second stage introduces eukaryotic genomes, known for their intron-exon structures and gene lengths that can span tens of kilobases~\citep{dalla2024nucleotide}. Alongside this transition, the model’s context window is enlarged from 8K to 100K base pairs, accommodating multiple co-expressed genes and regulatory elements. 


\section{Experiments}
\subsection{Pre-training}
\paragraph{Data.}
We adopt masked language modeling (MLM)~\citep{devlin2018bert} and character-level tokenization for all our DNA models. Following our \emph{Evolutionary training} strategy, we conduct pre-training in two phases:
Stage 1 (Short-Sequence Pre-training): We use prokaryotic (bacterial and archaeal) DNA data from the OpenGenome dataset introduced in~\cite{nguyen2024sequence}, training on sequences of length 8k to learn fundamental nucleic acid patterns in shorter contexts.
Stage 2 (Multi-species Post-training): We then continue pre-training on a multi-species collection, the  Multispecies dataset, presented in~\cite{dalla2024nucleotide}. The sequences in this stage can be as long as 100k, spanning archaebacteria, fungi, vertebrate genomes, and more. This step exposes the model to a rich spectrum of evolutionary signals, enabling it to handle much longer contexts and to capture the diverse structural intricacies of eukaryotic DNA.
Hence, we propose two main models, \texttt{TrinityMicroDNA} and \texttt{TrinityDNA}, each with 1 billion (1B) parameters. The former is trained solely on prokaryotic data, while the latter builds upon the former by post-training on eukaryotic data. By transitioning from bacterial genomes to longer eukaryotic genomes, our method achieves broad coverage of genomic features across scales. More details in Appendix~\ref{sec:app_pretrain}.

\paragraph{Scaling Laws.}
Figure~\ref{fig:scaling_law} illustrates three key aspects of our model's scaling behavior across parameter sizes, pre-training context lengths (8k, 30k, 100k). 
\emph{(Left)} We plot the compute-perplexity frontier against total FLOPs for several architectures, demonstrating that our approach consistently outperforms baseline methods (Transformer, Caduceus, EVO, and EVO2) at every compute level in different parameter sizes (6M to 1B). 
\emph{(Right)} We examine the effect of increasing context window sizes on a eukaryotic benchmark, finding a steady drop in perplexity when moving from 8k to 30k, and a further substantial improvement at 100k. 

\begin{table}[b!]
    \centering
    \begin{tabular}{lcc|cc}

    \toprule
    \textbf{Components} & \textbf{W/O} & \textbf{W} & \textbf{W/O} & \textbf{W} \\

    \midrule

    GRC  & 2.731 & 2.599~\tiny{\textcolor{teal}{(-0.132)}} & - & - \\

    GFM  & 2.599 & 2.534~\tiny{\textcolor{teal}{(-0.065)}} & - & - \\

    SMWA & 2.534 & 2.544~\tiny{\textcolor{red}{(+0.010)}} & 64.5 & 44.5~\tiny{\textcolor{teal}{(-31\%)}} \\

    \bottomrule

    \end{tabular}

    \caption{Effect of components on pre-training perplexity (left side) and computational cost (left side, TFLOPs).}

    \label{tab:groove_fusion_effect}

\end{table}


\begin{figure*}[t!]
    \centering
    \includegraphics[width=0.9\linewidth]{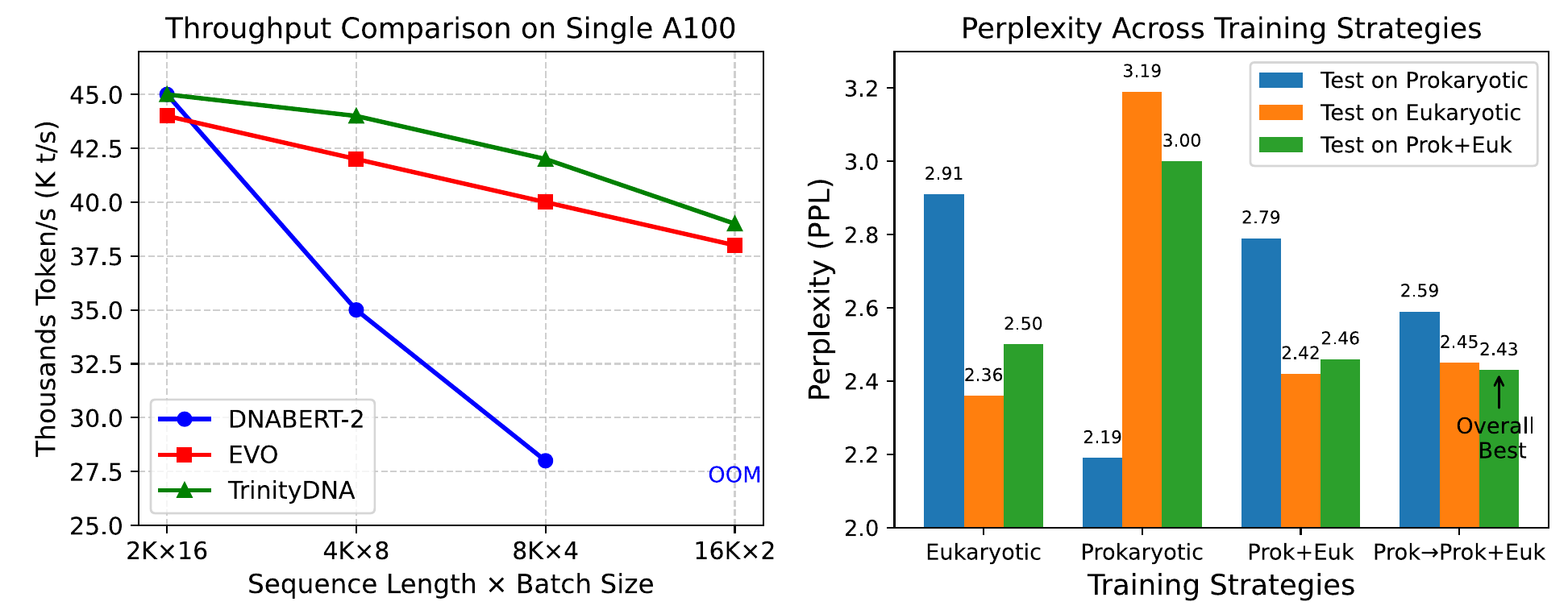}
    
    \caption{Comprehensive ablation study and efficiency analysis on TrinityDNA.}
    \label{fig:ablation}
\end{figure*}

\begin{table*}
\centering
\resizebox{0.85\linewidth}{!}{
\begin{tabular}{lcccccc}
\toprule
\textbf{Tasks} &
\textbf{DNABERT} &
\textbf{NT} &
\textbf{DNABERT2} &
\textbf{Caduceus} &
\textbf{HyenaDNA} &
\textbf{TrinityDNA} \\
\textbf{\# Params} & 86M & 2.5B & 117M & 40M & 1.6M & 1B \\ \midrule
H3 &
0.731 \tiny{$\pm0.015$} &
0.788 \tiny{$\pm0.012$} &
0.783 \tiny{$\pm0.014$} &
\underline{0.799 \tiny{$\pm0.029$}} &
0.779 \tiny{$\pm0.037$} &
\textbf{0.814 \tiny{$\pm0.014$}} \\
H3K14ac &
0.401 \tiny{$\pm0.018$} &
0.562 \tiny{$\pm0.015$} &
0.526 \tiny{$\pm0.019$} &
0.541 \tiny{$\pm0.212$} &
\underline{0.612 \tiny{$\pm0.065$}} &
\textbf{0.694 \tiny{$\pm0.016$}} \\
H3K36me3 &
0.473 \tiny{$\pm0.017$} &
\underline{0.620 \tiny{$\pm0.012$}} &
0.569 \tiny{$\pm0.015$} &
0.609 \tiny{$\pm0.109$} &
0.613 \tiny{$\pm0.041$} &
\textbf{0.692 \tiny{$\pm0.014$}} \\
H3K4me1 &
0.414 \tiny{$\pm0.012$} &
\underline{0.553 \tiny{$\pm0.011$}} &
0.505 \tiny{$\pm0.015$} &
0.488 \tiny{$\pm0.102$} &
0.512 \tiny{$\pm0.024$} &
\textbf{0.611 \tiny{$\pm0.015$}} \\
H3K4me2 &
0.323 \tiny{$\pm0.014$} &
0.365 \tiny{$\pm0.010$} &
0.311 \tiny{$\pm0.013$} &
0.388 \tiny{$\pm0.101$} &
\underline{0.455 \tiny{$\pm0.095$}} &
\textbf{0.480 \tiny{$\pm0.013$}} \\
H3K4me3 &
0.278 \tiny{$\pm0.015$} &
0.403 \tiny{$\pm0.011$} &
0.363 \tiny{$\pm0.014$} &
0.440 \tiny{$\pm0.202$} &
\underline{0.549 \tiny{$\pm0.056$}} &
\textbf{0.522 \tiny{$\pm0.015$}} \\
H3K79me3 &
0.612 \tiny{$\pm0.013$} &
0.647 \tiny{$\pm0.010$} &
0.674 \tiny{$\pm0.012$} &
\underline{0.676 \tiny{$\pm0.026$}} &
0.672 \tiny{$\pm0.048$} &
\textbf{0.741 \tiny{$\pm0.014$}} \\
H3K9ac &
0.512 \tiny{$\pm0.015$} &
0.560 \tiny{$\pm0.013$} &
0.556 \tiny{$\pm0.017$} &
\underline{0.604 \tiny{$\pm0.048$}} &
0.581 \tiny{$\pm0.061$} &
\textbf{0.662 \tiny{$\pm0.014$}} \\
H4 &
0.793 \tiny{$\pm0.012$} &
\underline{0.817 \tiny{$\pm0.015$}} &
0.807 \tiny{$\pm0.011$} &
0.789 \tiny{$\pm0.020$} &
0.763 \tiny{$\pm0.044$} &
\textbf{0.829 \tiny{$\pm0.012$}} \\
H4ac &
0.372 \tiny{$\pm0.016$} &
0.491 \tiny{$\pm0.018$} &
0.504 \tiny{$\pm0.019$} &
0.525 \tiny{$\pm0.240$} &
\underline{0.564 \tiny{$\pm0.038$}} &
\textbf{0.632 \tiny{$\pm0.013$}} \\
Human TF &
\underline{0.642 \tiny{$\pm0.012$}} &
0.633 \tiny{$\pm0.015$} &
0.701 \tiny{$\pm0.020$} &
- &
- &
\textbf{0.714 \tiny{$\pm0.009$}} \\
Mouse TF &
0.564 \tiny{$\pm0.018$} &
0.670 \tiny{$\pm0.014$} &
\underline{0.680 \tiny{$\pm0.015$}} &
- &
- &
\textbf{0.786 \tiny{$\pm0.012$}} \\
Promoter &
0.768 \tiny{$\pm0.015$} &
\underline{0.799 \tiny{$\pm0.012$}} &
0.774 \tiny{$\pm0.019$} &
- &
- &
\textbf{0.803 \tiny{$\pm0.012$}} \\
Splice Recon. &
0.841 \tiny{$\pm0.010$} &
\underline{0.894 \tiny{$\pm0.014$}} &
0.850 \tiny{$\pm0.020$} &
- &
- &
\textbf{0.927 \tiny{$\pm0.009$}} \\
Virus Covid &
0.555 \tiny{$\pm0.017$} &
\textbf{0.730 \tiny{$\pm0.012$}} &
\underline{0.710 \tiny{$\pm0.014$}} &
- &
- &
0.706 \tiny{$\pm0.015$} \\ \midrule
Overall Avg &
0.552 &
\underline{0.636} &
0.621 &
0.586 &
0.610 &
\textbf{0.708} \\
\bottomrule
\end{tabular}
}

\caption{GUE Benchmark performance comparison. Metrics are single/multi-label binary classification (MCC).}
\label{tab:gue_results}

\end{table*}

\begin{table*}[t]
\centering
\resizebox{0.85\linewidth}{!}{
    \begin{tabular}{llcccccc}
    \toprule
    \textbf{Task Type} & \textbf{Task} & \textbf{TrinityMicroDNA} & \textbf{TrinityDNA} & \textbf{EVO} & \textbf{EVO2} & \textbf{EVO2} & \textbf{Caduceus} \\
    \midrule
    & \# Params & 1B & 1B & 7B & 40B & 1B & 40M \\
    \midrule
    \multirow{7}{*}{\shortstack[l]{\textbf{RNA DMS}\\(\textit{Prokaryote})}} 
    & Zhang        & \textbf{0.560} & \underline{0.476} & 0.239 & 0.021 & 0.152 & 0.133 \\
    & Pitt         & \textbf{0.294} & 0.116 & 0.021 & 0.011 & 0.040 & \underline{0.175} \\
    & Hayden       & \textbf{0.365} & \underline{0.141} & 0.138 & 0.065 & 0.010 & 0.059 \\
    & Guy          & \underline{0.370} & 0.321 & 0.214 & \textbf{0.417} & 0.354 & 0.019 \\
    & Kobori       & \textbf{0.569} & \underline{0.561} & 0.255 & 0.226 & 0.317 & 0.275 \\
    & Domingo      & \underline{0.438} & 0.372 & \textbf{0.456} & 0.403 & 0.315 & 0.215 \\
    & Andreasson   & \textbf{0.292} & \underline{0.276} & 0.053 & 0.127 & 0.036 & 0.070 \\
    \midrule
    \multirow{6}{*}{\shortstack[l]{\textbf{Protein DMS}\\(\textit{Prokaryote})}} 
    & Firnberg     & \textbf{0.673} & 0.433 & 0.552 & 0.499 & \underline{0.621} & 0.018 \\
    & Jacquier     & \textbf{0.659} & 0.409 & 0.471 & 0.446 & \underline{0.529} & 0.023 \\
    & Kelsic       & \underline{0.406} & 0.397 & 0.321 & 0.309 & \textbf{0.463} & 0.047 \\
    & Weeks        & \textbf{0.573} & 0.505 & 0.526 & 0.492 & \underline{0.561} & 0.034 \\
    & Rockah       & 0.592 & 0.411 & 0.574 & \textbf{0.715} & \underline{0.657} & 0.168 \\
    & Chen         & 0.383 & 0.344 & 0.448 & \textbf{0.630} & \underline{0.534} & 0.053 \\
    \midrule
    \multirow{3}{*}{\textbf{DNA}} 
    & ClinVar               & 0.629 & \underline{0.933} & 0.555 & \textbf{0.950} & 0.927 & 0.657 \\
    & ClinVar-non coding    & 0.503 & \underline{0.931} & 0.397 & \textbf{0.974} & 0.920 & 0.601 \\
    & ClinVar-coding        & 0.654 & \textbf{0.930} & 0.484 & \underline{0.910} & 0.898 & 0.699 \\
    \midrule
    \multirow{3}{*}{\shortstack[l]{\textbf{Protein DMS}\\(\textit{Eukaryote})}} 
    & Sun          & 0.202 & \underline{0.315} & 0.314 & 0.295 & \textbf{0.334} & 0.097 \\
    & Tsuboyama    & 0.595 & 0.708 & 0.595 & \textbf{0.773} & \underline{0.717} & 0.134 \\
    & Faure        & 0.067 & \textbf{0.609} & 0.284 & 0.381 & \underline{0.482} & 0.041 \\
    \midrule
    \multicolumn{2}{l}{\textbf{Average Performance (Prokaryote)}} & \textbf{0.475} & \underline{0.366} & 0.328 & 0.335 & 0.353 & 0.099 \\
    \multicolumn{2}{l}{\textbf{Average Performance (Eukaryote)}}  & 0.404 & \textbf{0.699} & 0.415 & 0.667 & \underline{0.670} & 0.314 \\
    \bottomrule
    \end{tabular}
}
    \caption{Zero-shot performance across DNA, RNA, and protein DMS tasks. ClinVar tasks are binary classification (AUC); others are RNA/protein fitness regression (Spearman).}
\label{tab:zeroshot_results}

\end{table*}

\begin{table*}[ht]
\centering
\resizebox{0.85\linewidth}{!}{
    \begin{tabular}{llcccccc}
    \toprule
    \multirow{2}{*}{\textbf{Category}} & \multirow{2}{*}{\textbf{Models}} & \multicolumn{3}{c}{\textbf{Exact Match}} & \multicolumn{3}{c}{\textbf{75\% Match}} \\
    \cline{3-8}
    & & \textbf{Recall} & \textbf{Precision} & \textbf{F1} & \textbf{Recall} & \textbf{Precision} & \textbf{F1} \\ \midrule
    \multirow{4}{2cm}{\textbf{Pre-trained \\ Models}}
    & TrinityMicroDNA-1B   & 0.775 & \textbf{0.740} & \textbf{0.754} & 0.826 & \textbf{0.788} & 0.803 \\
    & TrinityMicroDNA-470M & 0.743 & 0.623 & 0.692 & 0.803 & 0.693 & 0.755 \\
    & TrinityMicroDNA-6M   & 0.592 & 0.333 & 0.488 & 0.723 & 0.445 & 0.524 \\
    & Caduceus-40M         & 0.149 & 0.148 & 0.140 & 0.194 & 0.189 & 0.180 \\ \midrule
    \multirow{3}{2cm}{\textbf{Classical \\ Pipelines}} 
    & Prodigal             & \textbf{0.832} & 0.666 & 0.725 & \textbf{0.909} & 0.765 & \textbf{0.829} \\
    & GENSCAN              & 0.721 & 0.681 & 0.702 & 0.810 & 0.774 & 0.799 \\
    & Glimmer              & 0.704 & 0.663 & 0.688 & 0.802 & 0.760 & 0.780 \\ \bottomrule
    \end{tabular}
}
    
    \caption{Results of CDS Annotation Benchmark on the filtered \textit{RefSeq} test set.}
\label{tab:gomc}

\end{table*}

%
%

\subsection{Ablation Study and Analysis}

\paragraph{(1) Effectiveness.} 
Table~\ref{tab:groove_fusion_effect} shows that GFM consistently lowers perplexity by modeling the spatial `groove' features of DNA sequences. 
Likewise, incorporating GRC (captures reverse-complement patterns) yields a notable drop in PPL, reflecting the importance of complementary-strand information. 
Meanwhile, SMWA enables multi-scale context handling, trading off some computational overhead for competitive perplexity across local and longer-range dependencies. Although SMWA incurs some performance loss, it is still evident that it saves resources for long sequence training.

\paragraph{(2) Efficiency.}
The left panel of Figure~\ref{fig:ablation} contrasts token-throughput as we sweep both sequence length and micro-batch size on 1B scale. Across every setting, TrinityDNA remains clearly at the top, retaining more than 80\% of its short-sequence throughput even at 64k tokens. This robustness stems from its sliding multi-window attention and optimized fused kernels, which maintain memory traffic at a nearly constant level as the context grows. 

\paragraph{(3) Training Strategies.}
The right plot in Figure~\ref{fig:ablation} shows a comprehensive ablation study separating the contributions of dataset size and evolutionary training strategy. The table shows perplexity scores across different model configurations. Key finding: Models initialized with weights from prokaryotic pre-training and then fine-tuned on combined data show better performance than models trained from scratch on the combined dataset. This validates both the importance of large, diverse datasets and our EST.

\subsection{Downstream Tasks}
We employ \texttt{TrinityMicroDNA} and \texttt{TrinityDNA} with 1B parameters as our base model for downstream evaluation, and we use LoRA tuning except for the zero-shot task.

\subsubsection{GUE Benchmark}
We evaluate our models on a comprehensive {Genomic Understanding Evaluation} (GUE) benchmark comprising tasks from \emph{Genomics Benchmark}~\citep{zhou2023dnabert} and \emph{Nucleotide Transformer} tasks~\citep{dalla2024nucleotide} by standard LoRA fine-tuning~\citep{hu2021lora}. For baselines, results are used in the original paper.

\paragraph{Results}
In Table~\ref{tab:gue_results}, we compare models DNABERT, Nucleotide Transformer (NT), DNABERT2, and \texttt{TrinityDNA}. 
Overall, ours outperforms prior methods across many metrics.
We observe large improvements in tasks that demand recognition of extended promoter regions or higher-order structural features, aligning with our architectural design for multi-window attention and GRC-based reverse complement awareness. 

\subsubsection{Zero-shot Performance}
We evaluated our two 1B models on 19 zero-shot downstream tasks—covering DNA pathogenicity (ClinVar), seven RNA DMS benchmarks and fifteen protein-fitness benchmarks across prokaryotic and eukaryotic (Table~\ref{tab:zeroshot_results}). The prokaryote-focused TrinityMicroDNA-1B dominates the prokaryotic regime, winning 8 of 13 prokaryotic tasks and attaining the highest prokaryotic average (0.475), whereas the multi-species TrinityDNA-1B excels on eukaryotic protein-fitness prediction, delivering the top score and the highest eukaryotic average (0.699), even surpassing the 40B EVO2 model (0.667). TrinityDNA also excels in state-of-the-art DNA pathogenicity performance, leading in CliVar-coding and ranking second overall. These complementary strengths underscore the benefit of ETS, and a UMAP projection of genome-level embeddings for ten representative clades (Appendix~\ref{app:zero}) reveals clear taxonomic clustering, indicating that both models learn rich species signals without fine-tuning.

\subsubsection{CDS Annotation Benchmark}
We also introduce a novel CDS Annotation Benchmark, aiming to assess long-sequence inference capabilities,
practical utility for gene annotation in real-world genomes.
From RefSeq, we collect all prokaryotic reference genomes and parse the GenBank annotation files for gene positions/types. This yields token-level labels indicating whether each token belongs to a coding sequence (CDS) with 20k sequence length and, if so, in which strand/direction it is transcribed. The detailed data statistics are described in Appendix~\ref{sec:app_cds}.

%

\paragraph{Results} The results are shown in Table~\ref{tab:gomc}. While Prodigal shows strong recall performance, TrinityMicroDNA-1B delivers the best precision and F1 scores for exact matches, highlighting the model's strong generalization capabilities across diverse datasets compared to classical pipelines.

\section{Conclusion}
We present \texttt{TrinityDNA}, a foundational DNA model that integrates biologically inspired modules—\emph{Groove Fusion} and \emph{Gated Reverse Complement}—alongside a \emph{multi-scale attention mechanism} to capture scales of genomic context effectively, built upon an ETS that transitions from prokaryotic to eukaryotic genomes. \texttt{TrinityDNA} gains strong generalization for diverse genomic prediction tasks. We also introduce the \emph{CDS Annotation Benchmark}, which evaluates coding sequence identification across organisms and provides a realistic standard for genome-scale annotation.

\bibliography{reference}

@inproceedings{gu2021efficiently,
  title     = {Efficiently Modeling Long Sequences with Structured State Spaces},
  author    = {Gu, Albert and Goel, Karan and R{\'e}, Christopher},
  booktitle = {International Conference on Learning Representations (ICLR)},
  year      = {2022},
  url       = {https://openreview.net/forum?id=uYLFoz1vlAC},
}

@misc{gu2021combining,
  title       = {Combining Recurrent, Convolutional, and Continuous-time Models with Linear State-Space Layers},
  author      = {Gu, Albert and Goel, Karan and R{\'e}, Christopher},
  howpublished= {arXiv preprint arXiv:2103.10897},
  year        = {2021},
  url         = {https://arxiv.org/abs/2103.10897},
}

@misc{gu2022parameterization,
  title       = {Parameterization and Initialization of Diagonal State Spaces for Sequence Modeling},
  author      = {Gu, Albert and Goel, Karan and R{\'e}, Christopher},
  howpublished= {arXiv preprint arXiv:2208.04933},
  year        = {2022},
  url         = {https://arxiv.org/abs/2208.04933},
}

@misc{gupta2022diagonal,
  title       = {Diagonally-Scalable Structured State Space Models for Long Sequence Modeling},
  author      = {Gupta, Hardik and Schuurmans, Dale and R{\'e}, Christopher},
  howpublished= {arXiv preprint arXiv:2211.07225},
  year        = {2022},
  url         = {https://arxiv.org/abs/2211.07225},
}

@misc{smith2022simplified,
  title       = {Simplified State Space Layers for Sequence Modeling},
  author      = {Smith, Samuel and Gu, Albert and Roberts, Daniel A. and R{\'e}, Christopher},
  howpublished= {arXiv preprint arXiv:2209.12951},
  year        = {2022},
  url         = {https://arxiv.org/abs/2209.12951},
}

@misc{dao2022hungry,
  title       = {Hungry Hungry Hippos: Towards Language Model Scaling via Backpropagation in Space},
  author      = {Dao, Tri and Jain, Shreyash and Rudra, Atri and R{\'e}, Christopher},
  howpublished= {arXiv preprint arXiv:2212.14052},
  year        = {2022},
  url         = {https://arxiv.org/abs/2212.14052},
}

@misc{poli2023hyena,
  title       = {{Hyena}: A Quasi-Attention Approach to Long-Range Language Modeling},
  author      = {Poli, Michael and Suteu, Iulian and others},
  howpublished= {arXiv preprint arXiv:2302.10866},
  year        = {2023},
  url         = {https://arxiv.org/abs/2302.10866},
}

@article{nguyen2024sequence,
   author = {Eric Nguyen and Michael Poli and Matthew G. Durrant and Brian Kang and Dhruva Katrekar and David B. Li and Liam J. Bartie and Armin W. Thomas and Samuel H. King and Garyk Brixi and Jeremy Sullivan and Madelena Y. Ng and Ashley Lewis and Aaron Lou and Stefano Ermon and Stephen A. Baccus and Tina Hernandez-Boussard and Christopher Ré and Patrick D. Hsu and Brian L. Hie },
   title = {Sequence modeling and design from molecular to genome scale with Evo},
   journal = {Science},
   volume = {386},
   number = {6723},
   pages = {eado9336},
   year = {2024},
   doi = {10.1126/science.ado9336},
   URL = {https://www.science.org/doi/abs/10.1126/science.ado9336},
}

@article{dalla2024nucleotide,
  title={Nucleotide Transformer: building and evaluating robust foundation models for human genomics},
  author={Dalla-Torre, Hugo and Gonzalez, Liam and Mendoza-Revilla, Javier and Lopez Carranza, Nicolas and Grzywaczewski, Adam Henryk and Oteri, Francesco and Dallago, Christian and Trop, Evan and de Almeida, Bernardo P and Sirelkhatim, Hassan and others},
  journal={Nature Methods},
  pages={1--11},
  year={2024},
  publisher={Nature Publishing Group US New York}
}

@article{zhou2023dnabert,
  title={Dnabert-2: Efficient foundation model and benchmark for multi-species genome},
  author={Zhou, Zhihan and Ji, Yanrong and Li, Weijian and Dutta, Pratik and Davuluri, Ramana and Liu, Han},
  journal={arXiv preprint arXiv:2306.15006},
  year={2023}
}

@article{dalla2023nucleotide,
  title={The nucleotide transformer: Building and evaluating robust foundation models for human genomics},
  author={Dalla-Torre, Hugo and Gonzalez, Liam and Mendoza-Revilla, Javier and Carranza, Nicolas Lopez and Grzywaczewski, Adam Henryk and Oteri, Francesco and Dallago, Christian and Trop, Evan and de Almeida, Bernardo P and Sirelkhatim, Hassan and others},
  journal={bioRxiv},
  pages={2023--01},
  year={2023},
  publisher={Cold Spring Harbor Laboratory}
}

@article{nguyen2023hyenadna,
  title={Hyenadna: Long-range genomic sequence modeling at single nucleotide resolution},
  author={Nguyen, Eric and Poli, Michael and Faizi, Marjan and Thomas, Armin and Birch-Sykes, Callum and Wornow, Michael and Patel, Aman and Rabideau, Clayton and Massaroli, Stefano and Bengio, Yoshua and others},
  journal={arXiv preprint arXiv:2306.15794},
  year={2023}
}

@article{ji2021dnabert,
  title={DNABERT: pre-trained Bidirectional Encoder Representations from Transformers model for DNA-language in genome},
  author={Ji, Yanrong and Zhou, Zhihan and Liu, Han and Davuluri, Ramana V},
  journal={Bioinformatics},
  volume={37},
  number={15},
  pages={2112--2120},
  year={2021},
  publisher={Oxford Academic}
}

@article{vaswani2017attention,
  title={Attention is all you need},
  author={Vaswani, Ashish and Shazeer, Noam and Parmar, Niki and Uszkoreit, Jakob and Jones, Llion and Gomez, Aidan N and Kaiser, {\L}ukasz and Polosukhin, Illia},
  journal={Advances in neural information processing systems},
  volume={30},
  year={2017}
}

@article{devlin2018bert,
  title={Bert: Pre-training of deep bidirectional transformers for language understanding},
  author={Devlin, Jacob and Chang, Ming-Wei and Lee, Kenton and Toutanova, Kristina},
  journal={arXiv preprint arXiv:1810.04805},
  year={2018}
}

@article{zaheer2020big,
  title={Big bird: Transformers for longer sequences},
  author={Zaheer, Manzil and Guruganesh, Guru and Dubey, Kumar Avinava and Ainslie, Joshua and Alberti, Chris and Ontanon, Santiago and Pham, Philip and Ravula, Anirudh and Wang, Qifan and Yang, Li and others},
  journal={Advances in neural information processing systems},
  volume={33},
  pages={17283--17297},
  year={2020}
}

@article{zhou2021towards,
  title={Towards a better understanding of reverse-complement equivariance for deep learning models in regulatory genomics},
  author={Zhou, Hannah and Shrikumar, Avanti and Kundaje, Anshul},
  journal={BioRxiv},
  pages={2020},
  year={2021}
}

@article{mallet2021reverse,
  title={Reverse-complement equivariant networks for DNA sequences},
  author={Mallet, Vincent and Vert, Jean-Philippe},
  journal={Advances in Neural Information Processing Systems},
  volume={34},
  pages={13511--13523},
  year={2021}
}

@article{achiam2023gpt,
  title={Gpt-4 technical report},
  author={Achiam, Josh and Adler, Steven and Agarwal, Sandhini and Ahmad, Lama and Akkaya, Ilge and Aleman, Florencia Leoni and Almeida, Diogo and Altenschmidt, Janko and Altman, Sam and Anadkat, Shyamal and others},
  journal={arXiv preprint arXiv:2303.08774},
  year={2023}
}

@article{team2023gemini,
  title={Gemini: a family of highly capable multimodal models},
  author={Team, Gemini and Anil, Rohan and Borgeaud, Sebastian and Wu, Yonghui and Alayrac, Jean-Baptiste and Yu, Jiahui and Soricut, Radu and Schalkwyk, Johan and Dai, Andrew M and Hauth, Anja and others},
  journal={arXiv preprint arXiv:2312.11805},
  year={2023}
}

@misc{liu2024chela,
      title={Short-Long Convolutions Help Hardware-Efficient Linear Attention to Focus on Long Sequences}, 
      author={Zicheng Liu and Siyuan Li and Li Wang and Zedong Wang and Yunfan Liu and Stan Z. Li},
      year={2024},
      eprint={2406.08128},
      archivePrefix={arXiv},
      primaryClass={cs.LG},
      url={https://arxiv.org/abs/2406.08128}, 
}

@misc{liu2024longvq,
      title={LongVQ: Long Sequence Modeling with Vector Quantization on Structured Memory}, 
      author={Zicheng Liu and Li Wang and Siyuan Li and Zedong Wang and Haitao Lin and Stan Z. Li},
      year={2024},
      eprint={2404.11163},
      archivePrefix={arXiv},
      primaryClass={cs.LG},
      url={https://arxiv.org/abs/2404.11163}, 
}

@inproceedings{icml2024vqdna,
    title={VQDNA: Unleashing the Power of Vector Quantization for Multi-Species Genomic Sequence Modeling},
    author={Siyuan Li and Zedong Wang and Zicheng Liu and Di Wu and Cheng Tan and Jiangbin Zheng and Yufei Huang and Stan Z. Li},
    booktitle={International Conference on Machine Learning (ICML)},
    year={2024}
}

@misc{liu2024genbench,
      title={GenBench: A Benchmarking Suite for Systematic Evaluation of Genomic Foundation Models}, 
      author={Zicheng Liu and Jiahui Li and Siyuan Li and Zelin Zang and Cheng Tan and Yufei Huang and Yajing Bai and Stan Z. Li},
      year={2024},
      eprint={2406.01627},
      archivePrefix={arXiv},
      primaryClass={q-bio.GN},
      url={https://arxiv.org/abs/2406.01627}, 
}

@misc{hu2021lora,
      title={LoRA: Low-Rank Adaptation of Large Language Models}, 
      author={Edward J. Hu and Yelong Shen and Phillip Wallis and Zeyuan Allen-Zhu and Yuanzhi Li and Shean Wang and Lu Wang and Weizhu Chen},
      year={2021},
      eprint={2106.09685},
      archivePrefix={arXiv},
      primaryClass={cs.CL},
      url={https://arxiv.org/abs/2106.09685}, 
}

@misc{dreos2013epdnew,
  title     = {EPD and EPDnew, high-quality promoter resources in the next-generation sequencing era},
  author    = {R. Dreos and G. Ambrosini and R. C. Perier and P. Bucher},
  year      = {2013},
  url       = {https://doi.org/10.1093/nar/gks1233},
  note      = {Nucleic Acids Research}
}

@misc{encode2012integrated,
  title     = {An integrated encyclopedia of DNA elements in the human genome},
  author    = {{The ENCODE Project Consortium}},
  year      = {2012},
  url       = {https://doi.org/10.1038/nature11247},
  note      = {Nature, 489(7414):57--74}
}

@misc{wang2019splice,
  title     = {SpliceFinder: Differential splicing detection using RNNs},
  author    = {R. Wang and X. Bai and K. Chen},
  year      = {2019},
  url       = {https://doi.org/10.1093/bioinformatics/btz092},
  note      = {Bioinformatics, 35(14):2373--2380}
}

@misc{benson2012genbank,
  title     = {GenBank},
  author    = {D. A. Benson and M. Cavanaugh and K. Clark and I. Karsch-Mizrachi and D. J. Lipman and J. Ostell and E. W. Sayers},
  year      = {2012},
  url       = {https://doi.org/10.1093/nar/gks1195},
  note      = {Nucleic Acids Research, 41(D1):D36--D42}
}

@misc{stamatoyannopoulos2012global,
  title     = {An encyclopedia of mouse DNA elements (Mouse ENCODE)},
  author    = {J. A. Stamatoyannopoulos and M. Snyder and R. Hardison and B. Ren and T. Gingeras and B. E. Bernstein and others},
  year      = {2012},
  url       = {https://doi.org/10.1186/gb-2012-13-8-418},
  note      = {Genome Biology, 13(8):418}
}

@misc{khare2021gisaid,
  title     = {GISAID’s role in pandemic response},
  author    = {S. Khare and C. Gurry and L. Freitas and B. B. Kelly and S. Maurer-Stroh and others},
  year      = {2021},
  url       = {https://doi.org/10.46234/ccdcw2021.255},
  note      = {China CDC Weekly, 3(49):1049--1051}
}

@misc{nebrao2021prokaryotic,
  title     = {No one tool to rule them all: prokaryotic gene prediction tool annotations are highly dependent on the organism of study},
  author    = {M. J. Nebrao and V. T. Kung and K. Rowe and G. R. Amaral and R. K. Azad and G. Cambray},
  year      = {2021},
  url       = {https://doi.org/10.1093/bioinformatics/btab827},
  note      = {Bioinformatics}
}

@misc{wang2024understand,
      title={Understanding and Mitigating Bottlenecks of State Space Models through the Lens of Recency and Over-smoothing}, 
      author={Peihao Wang and Ruisi Cai and Yuehao Wang and Jiajun Zhu and Pragya Srivastava and Zhangyang Wang and Pan Li},
      year={2024},
      eprint={2501.00658},
      archivePrefix={arXiv},
      primaryClass={cs.LG},
      url={https://arxiv.org/abs/2501.00658}, 
}

@misc{xu2006genome,
  title     = {Genome-wide average gene length is highly conserved but recombination rate varies among prokaryotes and eukaryotes},
  author    = {Xu, Z. and Jin, L.},
  year      = {2006},
  archivePrefix = {arXiv},
  primaryClass  = {q-bio.GN},
  url       = {https://academic.oup.com/mbe/article/23/6/1107/1055387}
}

@misc{schiff2024caduceus,
      title={Caduceus: Bi-Directional Equivariant Long-Range DNA Sequence Modeling}, 
      author={Yair Schiff and Chia-Hsiang Kao and Aaron Gokaslan and Tri Dao and Albert Gu and Volodymyr Kuleshov},
      year={2024},
      eprint={2403.03234},
      archivePrefix={arXiv},
      primaryClass={q-bio.GN},
      url={https://arxiv.org/abs/2403.03234}, 
}

@article{xiao2024duo,
        title={DuoAttention: Efficient Long-Context LLM Inference with Retrieval and Streaming Heads},
        author={Xiao, Guangxuan and Tang, Jiaming and Zuo, Jingwei and Guo, Junxian and Yang, Shang and Tang, Haotian and Fu, Yao and Han, Song},
        journal={arXiv},
        year={2024}
}

@article{notin2023proteingym,
  title={Proteingym: Large-scale benchmarks for protein fitness prediction and design},
  author={Notin, Pascal and Kollasch, Aaron and Ritter, Daniel and Van Niekerk, Lood and Paul, Steffanie and Spinner, Han and Rollins, Nathan and Shaw, Ada and Orenbuch, Rose and Weitzman, Ruben and others},
  journal={Advances in Neural Information Processing Systems},
  volume={36},
  pages={64331--64379},
  year={2023}
}

\newpage

\appendix

\newpage
\onecolumn
\section{Appendix of TrinityDNA}
\label{sec:app_pretrain}
\section{Experimental Setups}

In this appendix, we offer comprehensive details regarding the experimental setup for our research. 

\subsubsection{Hardware Configuration}
We conduct our experiments on a cluster of 31 host machines. Each machine is equipped with 8 A100 GPUs, providing substantial parallel computing power. The total number of CPU cores available across the cluster is 128, and the total memory amounts to 1007 GB. The operating system used is Ubuntu with the kernel version 5.4.0 - 139 - generic. Additionally, we utilize RDMA (Remote Direct Memory Access) technology to enhance data transfer efficiency. The DCGM (Data Center GPU Manager) drive version is 525.147.05.

\subsection{Details for Log Influential Score Analysis}
\label{subsec:appendix_fig2}

The analysis of \textbf{log influential scores} was designed to quantitatively measure how the influence of an input token on an output token diminishes with increasing distance between them. This helps to empirically visualize the locality bias of different model architectures.

\begin{enumerate}
    \item[1] \textbf{Models and Dataset:} We utilized the official pretrained checkpoints for the SSM-based model {Caduceus-Phage} and the full-attention model {DNABERT2}. The analysis was conducted on the human reference genome, {HG-38}.

    \item[2] \textbf{Sequence Sampling:} We randomly selected a contiguous sequence of {20,000 base pairs} from the HG-38 genome to serve as the input for the models.

    \item[3] \textbf{Influence Calculation:} The influential score is defined as the gradient of a model's output with respect to its input, specifically $|\partial y_t / \partial x_s|$. This value quantifies the extent to which a change in the input token at position $s$ affects the model's prediction at position $t$.
    \begin{itemize}
        \item To compute this, we performed a {forward pass} with the 20k-token sequence.
        \item We then used {backpropagation} to calculate the gradients of the output logits ($y_t$) with respect to the input embeddings ($x_s$) for a chosen target position $t$ near the end of the sequence.
        \item We computed the absolute value of these gradients and took their logarithm, yielding the {log influential score}, $\log |\partial y_t / \partial x_s|$.
    \end{itemize}

    \item[4] \textbf{Plotting:} The scores were plotted against the relative distance between the input and output tokens, $(t - s)$. The curve for each model represents the average scores over multiple runs with different randomly sampled sequences. The vertical dashed line labeled {``OOM at 16k"} indicates the point at which DNABERT2 failed to process the sequence due to an out-of-memory error, a known limitation of its quadratic attention mechanism.
\end{enumerate}

\subsection{Details for Average Attention Entropy Analysis}
\label{subsec:appendix_fig3}

The experiment on \textbf{average attention entropy} was conducted to demonstrate the ``oversmoothing" phenomenon in standard Transformer models, where attention distributions become increasingly uniform as sequence lengths grow.

\begin{enumerate}
    \item[1] \textbf{Model Architecture:} We constructed a {standard two-layer Transformer encoder} with full self-attention. The model was intentionally kept simple to isolate the effect of sequence length on the attention mechanism itself, without confounding factors from more complex architectures.

    \item[2] \textbf{Datasets and Training:} We trained two separate instances of this model from scratch:
    \begin{itemize}
        \item One on a large corpus of English text from \textbf{Wikipedia (Text Sequence)}.
        \item Another on genomic data from the \textbf{HG-38 reference genome (Nucleotide Sequence)}.
    \end{itemize}
    This dual-dataset approach allowed us to compare the oversmoothing effect across different data modalities.

    \item[3] \textbf{Entropy Calculation:} After training, the models were evaluated on sequences of varying lengths, from $2^{10}$ (1024) to $2^{14}$ (16384) tokens.
    \begin{itemize}
        \item For each input sequence, we extracted the attention probability distributions, $p$, from every attention head at every position.
        \item We then calculated the \textbf{Shannon entropy} for each distribution using the formula $H(p) = -\sum_{i} p_i \log_2 p_i$. A higher entropy value indicates a more uniform (or "smoother") distribution, while a lower entropy value suggests a more focused, spiky distribution.
        \item The final \textbf{"Avg. Attn Entropy"} value for each sequence length shown on the y-axis is the mean entropy calculated across all attention heads and all token positions in the sequence.
    \end{itemize}
    
    \item[4] \textbf{Plotting:} The chart plots the average attention entropy as a function of sequence length. The x-axis is on a logarithmic scale ($2^N$) to clearly show the trend across several orders of magnitude in sequence length.
\end{enumerate}

\subsubsection{Training Framework}
Our training framework is built upon the Megatron and DeepSpeed training frameworks, integrating FlashAttention to accelerate attention computation. It supports DeepSpeed-Ulysess for efficient memory management and large-scale model training, while also incorporating Pipeline Parallelism, Model Parallelism, and Data Parallelism (4D). This unique combination allows us to train models with longer context lengths and more complex architectures without compromising performance. Notably, while DeepSpeed-Ulysess currently does not support Pipeline Parallelism, our framework has successfully integrated both DeepSpeed-Ulysess and Pipeline Parallelism alongside Model and Data Parallelism, enabling a highly scalable and efficient 4D parallelism approach for state-of-the-art model training.
For our experiments, we build upon the PyTorch and PyTorch Lightning frameworks, which are widely recognized for their flexibility and efficiency in deep learning research. 
Our model adopts a Transformer encoder-only architecture. We incorporate several advanced techniques within this architecture:
\paragraph{Activation Function}: We use the GEGLU (Gated Gaussian Error Linear Unit) activation function, which has shown superior performance in handling complex data patterns compared to traditional activation functions.
\paragraph{Normalization Layers}: To stabilize the training process, we employ DeepNorm and LayerNorm. DeepNorm helps in mitigating the vanishing and exploding gradient problems, especially in deep neural networks, while LayerNorm normalizes the input across the feature dimension.
\paragraph{Positional Encoding}: We use RoPE (Rotary Positional Embedding) extended with Dynamic NTK scaling. RoPE is a powerful positional encoding method that can effectively capture the relative position information in sequences, and the Dynamic NTK scaling further enhances its ability to handle long-range dependencies.
\paragraph{Attention Mechanism}: We leverage Flash Attention 2, which significantly reduces the computational complexity of the attention mechanism, enabling faster and more memory-efficient training, especially for long sequences.

\subsubsection{Training Optimization}
We utilize DeepSpeed - Ulysses for training Transformer models with extremely long sequences. This framework is further modified and optimized based on the pipe mode. 

Regarding the training settings, we set the ZERO\_STAGE to 1, which helps in reducing memory usage during training by partitioning the optimizer states. We use the BF16 (Brain Floating Point 16) data type for the model parameters to speed up the training process, while the gradient accumulation data type is set to FP32 to maintain numerical stability.

Unless otherwise specified, we use cross-entropy loss as our objective function for training, which is a common choice for classification-related tasks in genomic sequence modeling. 

\begin{table}[htbp]
\centering
\caption{Model Configuration of Pretraining}
\resizebox{0.9\linewidth}{!}{
    \begin{tabular}{ccccccc}
    \toprule
     Size & Sequence Length & Layers & Hidden Size & FFN Hidden Size & Num Heads & Learning Rate \\
    \midrule
     6M & 8192 & 8 & 256 & 682 & 8 & 1.00E-03 \\
     40M & 8192 & 10 & 576 & 1536 & 8 & 6.00E-04 \\
     85M & 8192 & 12 & 768 & 2048 & 12 & 5.50E-04 \\
     170M & 8192 & 24 & 768 & 2048 & 16 & 5.00E-04 \\
     470M & 8192 & 24 & 1280 & 3413 & 20 & 4.00E-04 \\
     1B & 8192 & 24 & 2048 & 5461 & 32 & 3.00E-04 \\
     1B & 30720 & 24 & 2048 & 5461 & 32 & 2.00E-04 \\
     1B & 102400 & 24 & 2048 & 5461 & 32 & 1.00E-04 \\
    \bottomrule
    \end{tabular}
    }
\label{tab:pretrain_param}
\end{table}

\subsection{Data}

\paragraph{Data Sources}
The \emph{OpenGenome} dataset was composed by sampling bacterial and archaeal genomes from the GTDB v214.1, curated prokaryotic viruses from IMG/VR v4, and plasmid sequences from IMG/PR data databases.

\paragraph{Data Tokenization}
To facilitate model processing, we encode the input DNA sequences using token embedding vectors. Our vocabulary size is set to 5, consisting of the four nucleotide bases A, T, C, and G, and the special character N, which represents an unknown base. This encoding scheme transforms the biological sequences into a numerical format that can be readily processed by our model.

\paragraph{Pretraining Data Preparation}
During the pretraining phase, we adopt a masked language model (MLM) strategy. Specifically, we randomly select 15\% of the tokens in the DNA sequences as masked tokens. These selected tokens are then replaced with a special $<$mask$>$ token. This process allows the model to learn to predict the original nucleotides based on the context provided by the surrounding unmasked tokens.

However, since the $<$mask$>$ token is typically not present in downstream fine-tuning tasks, we introduce a more sophisticated replacement rule to mitigate the inconsistency between pretraining and fine-tuning. Among the 15\% of selected tokens:
1. With an 80\% probability, a token is replaced with the $<$mask$>$ token. To further align the long-sequence training process with downstream fine-tuning, there is a 0.02 probability that the replacement ratio can range from 0 to 80\%.
2. With a 10\% probability, a token is replaced with a randomly chosen token from the vocabulary.
3. With a 10\% probability, a token remains unchanged.

\paragraph{Sampling Strategy}
Inspiration for our sampling strategy is drawn from related works. Similar to using a single human reference genome and specific training and validation intervals in some studies, our approach is tailored to our integrated dataset. During training, we sample intervals from the integrated sequences and adjust the intervals at both ends to obtain sequences of length L. For the test set, we carefully select specific genomic regions to ensure the reliability of model evaluation. Although we do not follow the exact chromosome selection (chromosomes 14 and X) as in some references, we adopt a similar principle of using non-overlapping sequences of length L to evaluate the model's performance on unseen data. This sampling strategy helps to ensure that the model is trained on diverse and representative data and can generalize well to new sequences. 

\subsection{Models}
Table \ref{tab:pretrain_param} presents the model configurations for the pretraining phase, comparing the settings of baseline models and our proposed TrinityDNA model. 

\paragraph{Overall Configuration}
The table includes models with different parameter scales, denoted as 6M, 40M, 85M, 170M, 470M, and 1B. These models share several common structural features, such as the number of layers, hidden size, feed-forward network (FFN) hidden size, and the number of attention heads. The sequence length for most of the models is set to 8192, except for some TrinityDNA models, where the sequence lengths are 30720 and 102400. 

\paragraph{Baseline Models}
For the baseline models, as the model scale increases (from 6M to 1B), the number of layers, hidden size, FFN hidden size, and the number of attention heads generally increase. Correspondingly, the learning rate decreases, which is a common practice in deep learning to reset the training process for larger models. 

\paragraph{TrinityDNA Models}
The TrinityDNA models are designed to have the same basic configurations as the baselines for a fair comparison. For each model scale (e.g., 6M, 40B), the TrinityDNA model has identical hyperparameters to its corresponding baseline model in terms of sequence length, number of layers, hidden size, FFN hidden size, number of attention heads, and learning rate when the sequence length is 8192. 

However, when considering the longer sequence lengths of 30720 and 102400 for the 1B TrinityDNA model, the learning rate is further reduced. This adjustment is likely to ensure stable training when dealing with much longer sequences, as longer sequences can introduce more complex dependencies and potentially lead to training instability. 

\section{Downstream Task Details}
\label{sec:app_downstream}
In our study, apart from the Zero-shot benchmark, all other downstream tasks are carried out in the form of LoRA (Low-rank Adaptation) fine-tuning~\citep{hu2021lora}. Table \ref{tab:gue_param} presents the detailed LoRA fine-tuning parameters for a variety of tasks, which is essential for understanding the specific configurations and experimental setups of each task.
\subsection{LoRA Fine-tuning Parameters on GUE Benchmark~\citep{zhou2023dnabert}}

\begin{table}[t!]
    \centering
    \caption{LoRA Fine-tuning Parameters}
    \begin{tabular}{lcccc}
        \toprule
        Task & lr & LoRA rank & LoRA alpha & batch size \\
        \midrule
        tf\_0 & 0.001 & 4 & 8 & 16 \\
        tf\_1 & 0.0001 & 4 & 8 & 16 \\
        tf\_2 & 0.001 & 4 & 8 & 16 \\
        tf\_3 & 0.0001 & 4 & 8 & 16 \\
        tf\_4 & 0.001 & 4 & 8 & 16 \\
        mouse\_0 & 0.001 & 4 & 8 & 16 \\
        mouse\_1 & 0.0001 & 24 & 48 & 16 \\
        mouse\_2 & 0.0001 & 4 & 8 & 16 \\
        mouse\_3 & 0.001 & 4 & 8 & 16 \\
        mouse\_4 & 0.0001 & 4 & 8 & 16 \\
        core\_all & 0.0001 & 4 & 8 & 16 \\
        core\_notata & 0.0001 & 4 & 8 & 16 \\
        core\_tata & 0.0001 & 24 & 48 & 16 \\
        300\_all & 0.0001 & 24 & 48 & 16 \\
        300\_notata & 0.0001 & 8 & 16 & 16 \\
        300\_tata & 0.001 & 4 & 8 & 16 \\
        splice\_reconstructed & 0.0001 & 24 & 48 & 16 \\
        virus\_covid & 0.0001 & 4 & 8 & 16 \\
        H3 & 0.001 & 4 & 8 & 16 \\
        H3K14ac & 0.001 & 4 & 8 & 64 \\
        H3K36me3 & 0.0001 & 96 & 192 & 16 \\
        H3K4me1 & 0.001 & 4 & 8 & 64 \\
        H3K4me2 & 0.001 & 4 & 8 & 64 \\
        H3K4me3 & 0.0001 & 48 & 96 & 16 \\
        H3K79me3 & 0.0005 & 24 & 48 & 64 \\
        H3K9ac & 0.001 & 4 & 8 & 32 \\
        H4 & 0.0001 & 4 & 8 & 32 \\
        H4ac & 0.001 & 4 & 8 & 32 \\
        \bottomrule
    \end{tabular}
    \label{tab:gue_param}
\end{table}

\subsubsection{Task Descriptions}

\paragraph{Promoter detection (Human).}
This task focuses on detecting proximal promoter regions, the critical genomic sequences that initiate transcription. Because these segments host numerous key regulatory elements, precise identification is essential for advancing our understanding of gene regulatory mechanisms and recognizing the genomic basis of various diseases. Following \cite{dreos2013epdnew}, promoter sequences are split into TATA and non-TATA based on whether they contain a TATA box motif. For each subgroup, we extract the region from \(-249\) to \(+50\) base pairs around the transcription start site (TSS) to form the positive (promoter) class. For the negative (non-promoter) class, we randomly select equally sized sequences that (1) contain a TATA motif but lie outside promoter regions (TATA non-promoters), or (2) are formed by random substitutions (non-TATA non-promoters). We also merge both TATA and non-TATA subsets into a combined dataset labeled \textit{all}.

\paragraph{Core promoter detection (Human).}
Similar to the proximal promoter task, this variant targets an even smaller window -34 to +35 base pairs around the TSS to capture the \emph{core promoter} region. Predicting the core region is more challenging due to the limited context, as it concentrates on the immediate surroundings of the TSS and the start codon.

\paragraph{Transcription factor binding site prediction (Human).}
This task involves forecasting transcription factor (TF) binding sites in the human genome. TF binding sites are central to gene expression regulation; accurate prediction helps decode intricate gene regulatory networks and identify therapeutic targets. We use 690 ENCODE ChIP-seq experiments~\cite{encode2012integrated}, covering 161 TF binding profiles in 91 cell lines. A 101-bp window around the center of each peak defines the positive (TFBS) class, while segments of the same length and matched GC content form the negative (non-TFBS) class. To avoid trivial or overly difficult tasks (e.g., F1 scores $>$ 0.95 or $<$ 0.50), we filter out such datasets and randomly pick 5 from the remaining pool.

\paragraph{Splice site prediction (Human).}
This task locates splice donor and acceptor sites in the human genome. Correct splice site identification is crucial for understanding protein diversity and the pathological effects of aberrant splicing. We use a dataset~\citep{wang2019splice} of 400-bp sequences from the Ensembl GRCh38 reference genome. Following \cite{ji2021dnabert}, we note that models often attain near-perfect results on the original set (10,000 splice donors, acceptors, and non-splice sites), which does not reflect the real difficulty of detecting non-canonical sites. To address this, we iteratively enrich the dataset with adversarial examples (previously unseen false positives) in a hold-out set, making the task substantially harder.

\paragraph{Enhancer--promoter interaction (Human).}
This binary classification task aims to identify whether an enhancer interacts with a promoter, a crucial relationship in the human genome that modulates gene expression. The input comprises sequence pairs (enhancer and promoter), and the output is a binary prediction indicating an interaction or lack thereof.

\paragraph{Species classification (Virus \& Fungi).}
Here, the goal is to classify species based on genomic segments. We build these datasets from viral and fungal reference genomes obtained from GenBank~\citep{benson2012genbank}.

\paragraph{Transcription factor binding site prediction (Mouse).}
Analogous to the human TFBS task, this variant deals with mouse genomes using Mouse ENCODE ChIP-seq data \cite{stamatoyannopoulos2012global}. We generate negative sequences by applying dinucleotide shuffling while preserving frequency. All other settings remain consistent with the human TFBS dataset. As with the human version, we filter out tasks with extreme F1 scores and randomly sample 5 datasets from the remaining collection.

\paragraph{Epigenetic marks prediction (Yeast).}
This task forecasts epigenetic modifications (e.g., DNA or histone modifications) in yeast. Such marks affect gene regulation without altering the DNA sequence. We downloaded 10 datasets from the following link: \url{http://www.jaist.ac.jp/tran/nucleosome/members.htm} and split each dataset into training, validation, and test sets at an 8:1:1 ratio.

\paragraph{Covid variant prediction (Virus).}
Focusing on SARS-CoV-2, this task predicts the virus's variant type from 1000-bp genomic fragments. We gather data from EpiCoV \cite{khare2021gisaid} provided by GISAID, covering 9 variants: Alpha, Beta, Delta, Eta, Gamma, Iota, Kappa, Lambda, and Zeta.

\paragraph{Configurations.}
As shown in Tab~\ref{tab:gue_param}, LoRA Rank and LoRA Alpha LoRA is a method used to fine-tune large pre-trained models more efficiently. The LoRA rank defines the rank of the low-rank matrices employed in the adaptation process, and LoRA alpha serves as a scaling factor. These two parameters jointly control the complexity and expressiveness of the LoRA adaptation. In the table, the LoRA rank ranges from 4 to 96, and the LoRA alpha ranges from 8 to 192. Different tasks use distinct combinations of LoRA rank and LoRA alpha, suggesting that the optimal LoRA configuration depends on the characteristics of each task.

\section{Zero-shot Details}
\label{app:zero}
\subsubsection{Datasets.}
\paragraph{DNA pathogenicity (ClinVar).}
We follow the pipeline of \citep{nguyen2024sequence}: single-nucleotide polymorphisms (SNPs) labelled as \emph{pathogenic} or \emph{benign} are taken from the ClinVar release, and up to 4k nucleotide flanking windows are retrieved from the GRCh38 human reference genome to form input sequences.

\paragraph{RNA deep-mutational scanning (DMS).}
Seven non-coding RNA DMS benchmarks—\emph{Zhang}, \emph{Pitt}, \emph{Hayden}, \emph{Guy}, \emph{Kobori}, \emph{Domingo} and \emph{Andreasson}—are collected exactly as in \citep{nguyen2024sequence}.  They cover diverse prokaryotic RNAs.

\paragraph{Protein DMS.}
Starting from the \textsc{proteinGYM} suite~\cite{notin2023proteingym}, we keep every experiment for which the wild-type codon context can be unambiguously reconstructed.
This yields nine prokaryotic sets (\emph{Firnberg}, \emph{Jacquier}, \emph{Kelsic}, \emph{Weeks}, \emph{Rockah}, \emph{Chen}) and one human set (\emph{Sun}).
To broaden the eukaryotic coverage, we further add two recently released human protein scans.
For cases where the original authors reported only amino-acid sequences, we rebuild nucleic-acid inputs by selecting the most frequent codon for each amino-acid according to a codon-usage table computed from all CDS regions of GRCh38.


\begin{figure*}[t!]
    
    \centering
    \includegraphics[width=\linewidth]{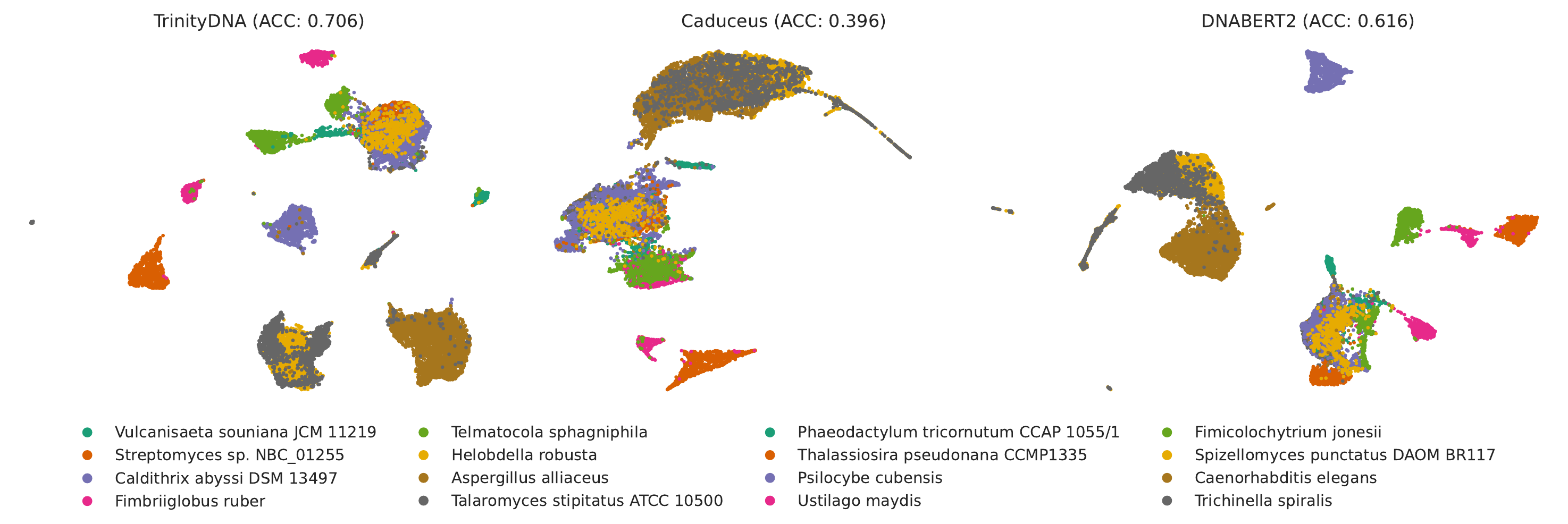}
    
    \caption{Zero-shot Embedding visualization of different species.}
    \label{fig:embedding}
\end{figure*}

\subsubsection{Zero-shot Metric Computation.}
For MLM-based models, we employ two metrics:
(1) \textit{Mean PPL over masked positions}: we sequentially mask token(s) and measure the resulting perplexity.
(2) \textit{Masked Mutation Impact}: We compare the PPL difference between wild-type (wt) and mutant (mt) sequences, focusing specifically on masked variant positions.
For generative models such as EVO, we rely on likelihood-based scores under the same masking scheme, assessing how well the model anticipates mutations.

\begin{figure*}[b!]
    \centering
    \includegraphics[width=0.8\linewidth]{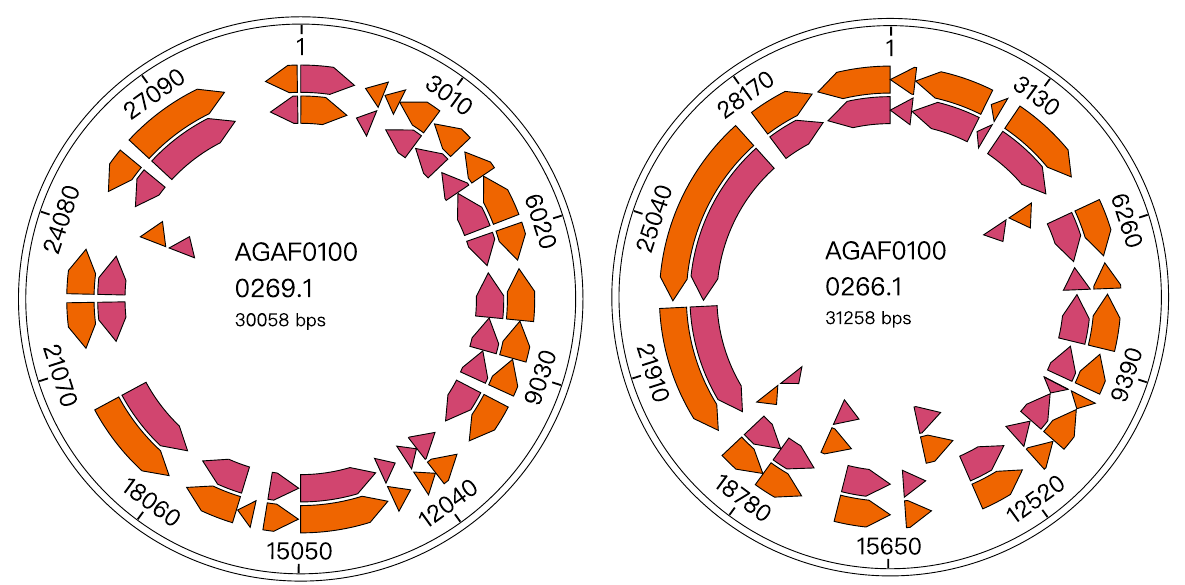}
    \caption{Comparison diagrams of two examples in the CDS annotation task. The orange parts represent the annotation results of Prodigal (serving as the ground truth), and the red parts represent the prediction results of TrinityDNA.}
    \label{fig:CDS_data}
\end{figure*}

\subsection{CDS Benchmark}
\label{sec:app_cds}

\subsubsection{Data and Configuration}

\begin{figure}
    \centering
    \includegraphics[width=0.95\linewidth]{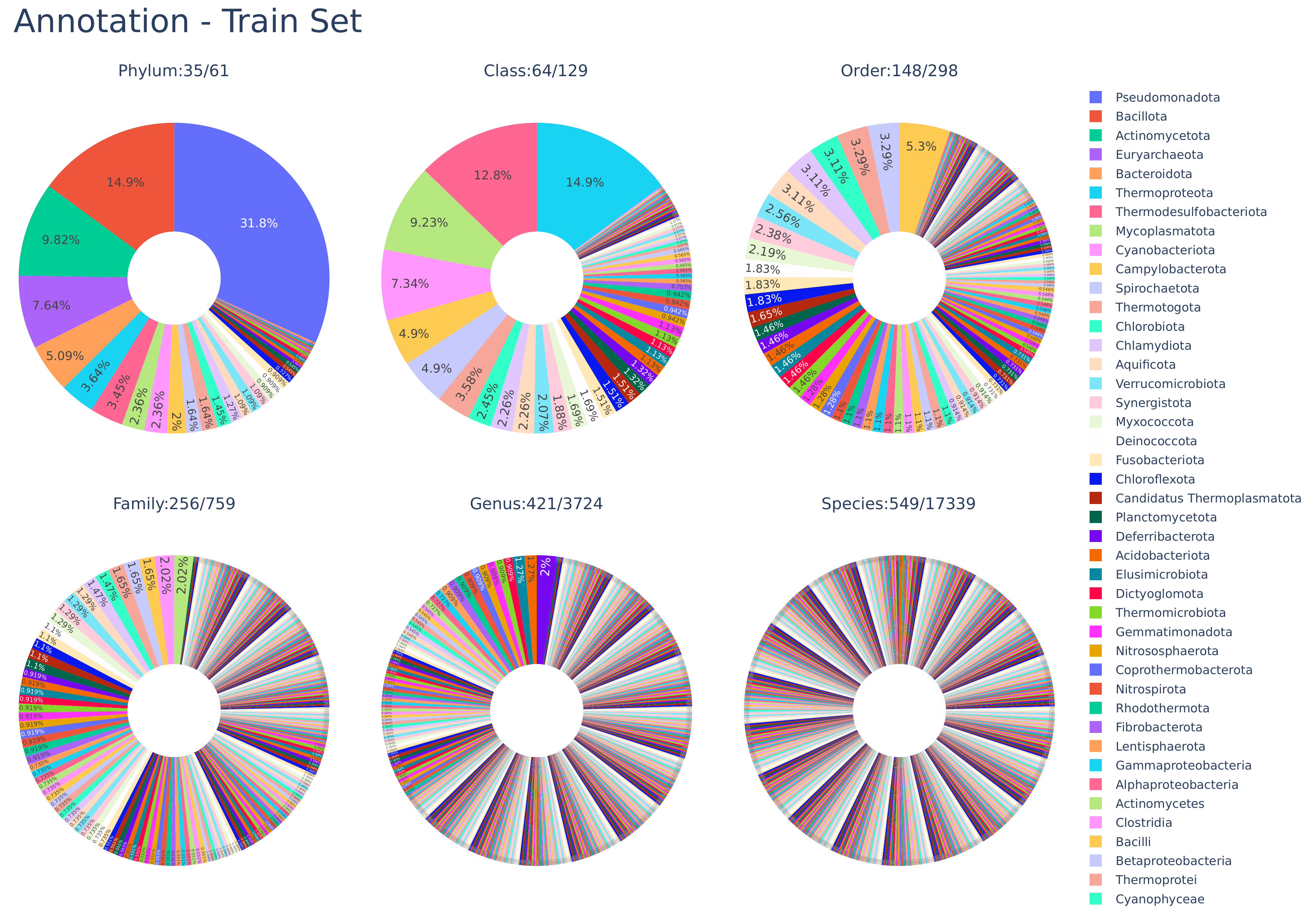}
    
    \caption{Distribution of taxonomic groups in the training and IID test sets. Our benchmark datasets are constructed from a comprehensive collection of bacterial and archaeal genomes (17,339 genomes across 61 phyla). From these, we selected 35 phyla for training, and the IID test set is randomly sampled from these training phyla, maintaining the same taxonomic distribution. This design enables evaluation of the model's ability to generalize within known taxonomic groups.}
    \label{fig:CDS_train}
    
\end{figure}

\begin{figure}
    \centering
    \includegraphics[width=0.95\linewidth]{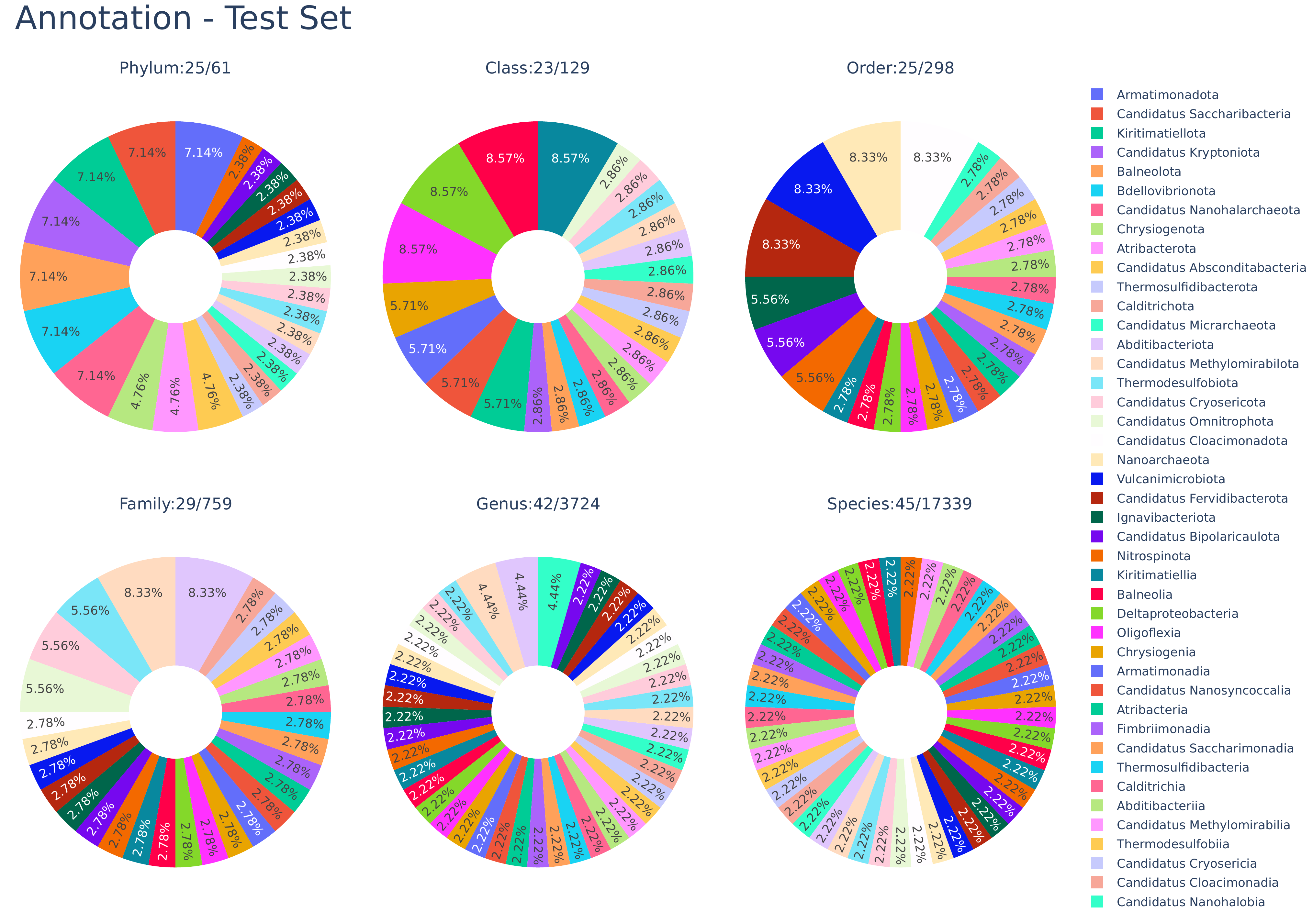}
    
    \caption{Distribution of taxonomic groups in the OOD test sets. To evaluate model generalization beyond the training distribution, we constructed two distinct OOD test sets: A phylogenetically diverse set of 45 genomes carefully selected from the remaining phyla not used in training, ensuring maximum coverage of untrained phyla.}
    \label{fig:CDS_test}
\end{figure}

\paragraph{Data Source.} Obtain all prokaryotic reference genomes from RefSeq, and construct token-level annotation labels based on gene localization and type information in GenBank~\citep{benson2012genbank} annotation files. 

\paragraph{Data Splits.}
\textit{RefSeq Dataset}: We exclude the 35 training phyla, leaving 26 additional phyla in RefSeq. We randomly sample two species per phylum (or one if only one species exists), yielding 45 genomes total. These serve as an out-of-distribution test set.
This benchmark tests both the model’s ability to handle very long sequences and its competence in gene structure recognition. By comparing performance on IID, metagenomic, and cross-phyla sets, we gauge the model’s robustness and generalization in realistic microbial genome annotation scenarios. The training and test data statistics are shown in Figure~\ref{fig:CDS_train} and \ref{fig:CDS_test}.

\paragraph{LoRA Tuning Setup}
We adopt Low-Rank Adaptation (LoRA)~\citep{hu2021lora} to fine-tune our model on the CDS Benchmark. Specifically, we use a learning rate of $r=0.001$ and a batch size of 32. For LoRA hyperparameters, we set the low-rank dimension $\text{LoRA-r}=4$ and the scaling factor $\text{LoRA-}\alpha=32$. Furthermore, we apply LoRA to two additional MLP layers in the classification head, whose hidden dimensions are 256 and 128, respectively. By decoupling the base model parameters from the low-rank adaptation parameters, LoRA allows us to retain the expressive capacity of the underlying large model while conferring adaptability to new domains with minimal additional parameters.

\subsubsection{Evaluation Metrics on the CDS Benchmark}
We evaluate gene annotations on the CDS Benchmark using standard metrics of recall, precision, and F\textsubscript{1}, where
\[
\text{Recall} = \frac{\text{TP}}{\text{TP} + \text{FN}}, 
\quad
\text{Precision} = \frac{\text{TP}}{\text{TP} + \text{FP}}, 
\quad
F_1 = 2 \times \frac{\text{Precision} \times \text{Recall}}{\text{Precision} + \text{Recall}}.
\]
Here, TP, FP, and FN represent true positives, false positives, and false negatives, respectively.

\paragraph{Exact Match.} 
A predicted coding sequence (CDS) is counted as a TP under \emph{Exact Match} if the predicted region’s start and end coordinates exactly match those of the reference annotation. In other words, the tool must recover the full CDS boundaries without offset.

\paragraph{75\% Match.}
In line with \citep{nebrao2021prokaryotic}, we introduce a less stringent matching criterion. Under \emph{75\% Match}, a predicted CDS is counted as a TP if:
\begin{enumerate}
  \item The predicted CDS lies fully within the boundaries of the true (annotated) CDS.
  \item The predicted CDS length is at least 75\% of the length of the true CDS.
  \item The predicted direction is matched.
\end{enumerate}
This relaxes the exact requirement on start/end positions and instead focuses on substantially overlapping the annotated region.



\section{Related Work}
\label{app:related}
\paragraph{Long-Sequence Model}
Transformers, with their multi-head attention, can capture such dependencies but suffer from quadratic computational complexity with sequence length, limiting their use in long-sequence tasks like genomic analysis~\citep{devlin2018bert}.
To address this, models like BigBird use sparse attention to expand context size~\citep{zaheer2020big}. Based on SSM-derived operators, a line of long-sequence models is proposed to handle long DNA sequences (up to 1M bps) but is unidirectional and not robust to RC inputs~\citep{liu2024chela, poli2023hyena, liu2024longvq, nguyen2024sequence}.

\paragraph{DNA Foundation Model}
Early DNA language models such as DNABERT were restricted by Transformer's complexity, with limited context sizes for DNA sequences~\citep{ji2021dnabert, zhou2023dnabert, dalla2023nucleotide,nguyen2023hyenadna,icml2024vqdna}.
Recent research focuses on integrating biological features. Caduceus (MambaDNA) exploits DNA's reverse-complement symmetry for better gene regulation modeling~\citep{mallet2021reverse}. Our TrinityDNA model innovatively adds the Groove Fusion module to capture DNA structural features.
In training, traditional models on single-type genomic data have limited generalization. Our evolutionary training strategy, with staged training on different genomes and increasing context lengths, enhances model adaptability and performance.

\section*{Impact Statements}
The work presented in this paper introduces TrinityDNA, a novel deep-learning model designed to address key challenges in DNA sequence modeling. By incorporating biologically informed components such as Groove Fusion and Gated Reverse Complement (GRC), along with a multi-scale attention mechanism and evolutionary training strategy, TrinityDNA makes significant strides in improving the accuracy and efficiency of genomic sequence analysis.

The impact of TrinityDNA extends beyond academic research in computational genomics. The model's ability to efficiently capture long-range dependencies in DNA sequences enables more accurate predictions of gene functions, regulatory mechanisms, and other biological insights. This can have profound implications for various fields, including personalized medicine, where understanding genetic variations is crucial for disease prevention and treatment. In addition, by improving the understanding of complex biological systems, TrinityDNA could accelerate advancements in biotechnology and drug development, where genomic data plays a pivotal role in identifying novel therapeutic targets and biomarkers.

Furthermore, the model's scalability and integration of multi-species genomic data position it as a tool with the potential to advance our understanding of the genomic underpinnings of diverse organisms. This has wide-reaching implications for evolutionary biology, conservation efforts, and the study of microbiomes, where large-scale comparative genomic analysis is essential. By bridging the gap between computational methods and biological insights, TrinityDNA serves as a significant step toward more robust, data-driven approaches to understanding the complexities of life at the genomic level.

\end{document}